\documentclass[%
 reprint,
 amsmath,amssymb,
 aps, nofootinbib
]{revtex4-1}
\usepackage{graphicx}
\usepackage{dcolumn}
\usepackage{bm}

\usepackage{amsmath,amssymb}
\usepackage{slashed}
\usepackage{graphicx}
\usepackage{dsfont}
\usepackage{bm}
\usepackage[top=1.0in,bottom=1.0in,left=1.0in,right=1.0in]{geometry}
\usepackage[colorlinks,linkcolor=blue,citecolor=blue]{hyperref}
\usepackage{CJKutf8}
\usepackage[caption=false]{subfig}

\begin{document}
\setlength{\oddsidemargin}{0.5cm}
\setlength{\topmargin}{-0.1cm}
\setlength{\textheight}{21cm}
\setlength{\textwidth}{15cm}
\newcommand{\be}{\begin{equation}}
\newcommand{\ee}{\end{equation}}
\newcommand{\bea}{\begin{eqnarray}}
\newcommand{\eea}{\end{eqnarray}}
\newcommand{\ba}{\begin{eqnarray}}
\newcommand{\ea}{\end{eqnarray}}

\newcommand{\fslash}{\hspace{-1.4ex}/\hspace{0.6ex} }
\newcommand{\Dslash}{D\hspace{-1.6ex}/\hspace{0.6ex} }
\newcommand{\Wslash}{W\hspace{-1.6ex}/\hspace{0.6ex} }
\newcommand{\pslash}{p\hspace{-1.ex}/\hspace{0.6ex} }
\newcommand{\kslash}{k\hspace{-1.ex}/\hspace{0.6ex} }
\newcommand{\underkslash}{{\underline k}\hspace{-1.ex}/\hspace{0.6ex} }
\newcommand{\epslash}{{\epsilon\hspace{-1.ex}/\hspace{0.6ex}}}
\newcommand{\partslash}{\partial\hspace{-1.6ex}/\hspace{0.6ex} }

\newcommand{\nn}{\nonumber}
\newcommand{\Tr}{\mbox{Tr}\;}
\newcommand{\tr}{\mbox{tr}\;}
\newcommand{\ket}[1]{\left|#1\right\rangle}
\newcommand{\bra}[1]{\left\langle#1\right|}
\newcommand{\rhoraket}[3]{\langle#1|#2|#3\rangle}
\newcommand{\brkt}[2]{\langle#1|#2\rangle}
\newcommand{\pdif}[2]{\frac{\partial #1}{\partial #2}}
\newcommand{\pndif}[3]{\frac{\partial^#1 #2}{\partial #3^#1}}
\newcommand{\pbm}[1]{\protect{\bm{#1}}}
\newcommand{\avg}[1]{\left\langle #1\right\rangle}
\newcommand{\vnabla}{\mathbf{\nabla}}
\newcommand{\notes}[1]{\fbox{\parbox{\columnwidth}{#1}}}
\newcommand{\pair}{\raisebox{-7pt}{\includegraphics[height=20pt]{pair0.pdf}}}
\newcommand{\paircrs}{\raisebox{-7pt}{\includegraphics[height=20pt]{pair0cross.pdf}}}
\newcommand{\paircc}{\raisebox{-7pt}{\includegraphics[height=20pt]{pair0cc.pdf}}}
\newcommand{\paircrscc}{\raisebox{-7pt}{\includegraphics[height=20pt]{pair0crosscc.pdf}}}
\newcommand{\pairloop}{\raisebox{-7pt}{\includegraphics[height=20pt]{pairloop.pdf}}}
\newcommand{\pairloopf}{\raisebox{-7pt}{\includegraphics[height=20pt]{pairloop4.pdf}}}
\newcommand{\pairlooph}{\raisebox{-7pt}{\includegraphics[height=20pt]{pair2looph.pdf}}}


\title{Tomography of Pions and Kaons in the QCD Vacuum:\\ Transverse Momentum Dependent Parton Distribution Functions}

\author{Wei-Yang Liu}
\email{wei-yang.liu@stonybrook.edu}
\affiliation{Center for Nuclear Theory, Department of Physics and Astronomy, Stony Brook University, Stony Brook, New York 11794-3800, USA}

\author{Ismail Zahed }
\email{ismail.zahed@stonybrook.edu}
\affiliation{Center for Nuclear Theory, Department of Physics and Astronomy, Stony Brook University, Stony Brook, New York 11794-3800, USA}

\date{\today}
\begin{abstract}
We evaluate the pion and kaon transverse momentum dependent parton distribution functions (TMDPDFs) in the instanton liquid model (ILM), a model of the QCD vacuum at low resolution. The relevant TMDs are 
factored into a constituent quark distribution times a rapidity dependent soft factor from staple-shaped Wilson lines, for fixed parton longitudinal
momentum and transverse separation. 
The results are evolved to higher
rapidities using the Collins-Soper (CS) kernel and higher resolution using renormalization
group evolution.  
The comparison to existing extractions of pion TMDs from Drell-Yan (DY) data is briefly discussed.
\end{abstract}

\maketitle

\section{Introduction}
Fast moving hadrons carry an increasing number of sub-constituents quarks and gluons in QCD, deemed partons. The parton distribution functions (PDFs) capture their longitudinal momentum distributions, the simplest of all partonic distributions in a hadron~\cite{Blumlein:2012bf}. PDFs play a central role in the description of inclusive and semi-inclusive
processes in high energy scattering, thanks to factorization. Transverse momentum distributions (TMDs)
allows for a spatial description of the partons in a fast moving hadron, by recording both the longitudinal momentum and transverse location of a given parton~\cite{Boussarie:2023izj}.

Transverse momentum dependent parton distribution functions (TMDPDFs)  play a central role in the analyses of a wide range of high energy data both at electron and hadron facilities worldwide. Their understanding is part of a large experimental effort at COMPASS at CERN, JLAB in
Virginia, LHC at CERN and the future EIC at BNL. Currently, TMDPDFs are empirically extracted from 
Drell-Yan (DY) processes and semi-inclusive deep inelastic scattering (SIDIS), with small final hadron momentum transfer~\cite{Rogers:2015sqa}. 

TMDPDFs, or beam functions, are defined as correlation functions of bilocal quarks connected with staple-like Wilson lines, which are inherently non-perturbative. Their unnderstanding from first principles has been challenging. Recent lattice developments using the large momentum effective theory (LaMET) have proven useful for their possible Euclidean extraction and matching using quasi-TMDs, where the gluonic soft function can be related to a mesonic form factor~\cite{LatticeParton:2020uhz}.

\begin{figure}
  \centering
    \includegraphics[scale=0.5]{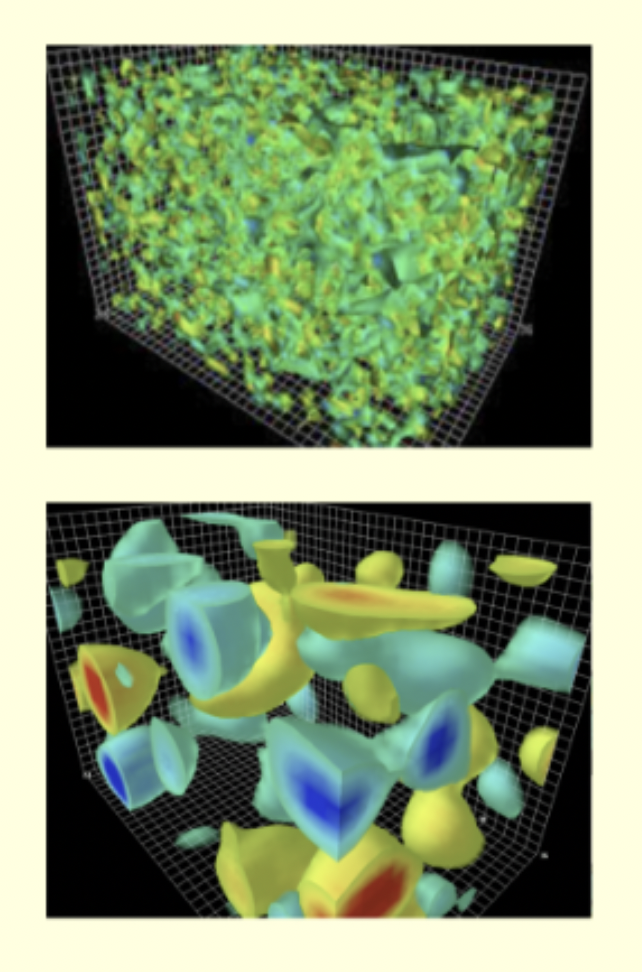}
    \caption{Visualization of the vacuum in gluodynamics, before cooling at a resolution of about $\frac 1{10}\,{\rm fm}$ (top),  and after cooling  at a resolution of about $ \frac 13\,{\rm fm}$ (bottom)~\cite{Moran:2008xq}, where the pseudoparticles emerge.}
    \label{fig:vacuum}
\end{figure}

Lattice QCD offers a first principle approach to address all  non-perturbative aspects of the theory, yet it does not provide the necessary insights to the physical mechanisms at work. In the continuum, these
mechanisms require a non-perturbative description of the vacuum state. The QCD vacuum as captured by the QCD instanton liquid model (ILM) offers by far the most compelling description of the gauge configurations 
at low resolution. In Fig.~\ref{fig:vacuum} we show
a snapshot of the vacuum state in gluodynamics at a high resolution of $\frac 1{10}\,\rm fm$ (top) and a low resolution of about $\frac 13\, \rm fm$ (bottom), following from a gradient flow technique~\cite{Michael:1994uu,Michael:1995br,Leinweber:1999cw,Bakas:2010by,Biddle:2018bst,Hasenfratz:2019hpg,Athenodorou:2018jwu,Biddle:2020eec}. 
The emergent ILM at low resolution is characterized by 
a density of pseudoparticles (instantons plus anti-instantons) $n=1\, \rm fm^{-4}$ and a typical size of
$\rho=\frac 13\,\rm fm$~\cite{Shuryak:1981ff}. Thanks to the low  packing fraction $\kappa=\pi^2\rho^4n\sim 0.1$, the ILM
is sufficiently dilute to allow for a many-body analysis 
using semi-classical techniques~\cite{Schafer:1996wv, Diakonov:1985eg,Liu:2023fpj,Liu:2023yuj,Shuryak:2022wtk,Shuryak:2022thi,Shuryak:2021hng,Shuryak:2021fsu,Liu:2024rdm,Liu:2024jno,Zahed:2022wae,Zahed:2021fxk} (and references therein).

The purpose of this paper is to develop an understanding of the TMDPDFs for the pion and kaon at low resolution, using the ILM. In section~\ref{SECII} we outline  the general aspects of pion TMDs, with those for the kaons following similarly. We suggest that in the ILM, the pion and kaon TMDs split into a bare constituent quark TMD distribution times a rapidity dependent soft factor capturing the contribution from Wilson staple. The constituent quark TMD distributions of pion and kaon 
are evaluated with the leading Fock state. 
In section~\ref{SECIII} we make use of the 
Collins-Soper-Sterman (CSS) renormalization group to evolve the TMDs to larger resolution. The results for the pion and kaon TMDs at low and high resolution are presented in section~\ref{SECIV}, with some comparison to existing DY data. Our conclusions are in section~\ref{SECV}. A number of Appendices are added to detail and support our arguments in the main text.

\begin{figure}
    \centering
\subfloat[\label{fig:wilson_1}]{\includegraphics[height=2cm,width=0.45\linewidth]{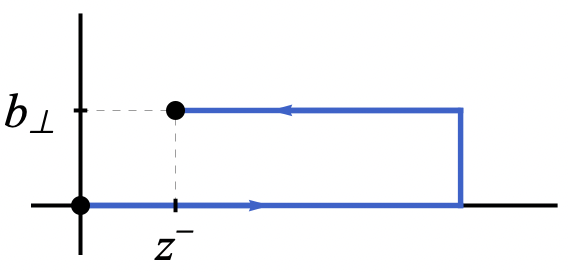}}
\subfloat[\label{fig:wilson_2}]{\includegraphics[height=2cm,width=0.54\linewidth]{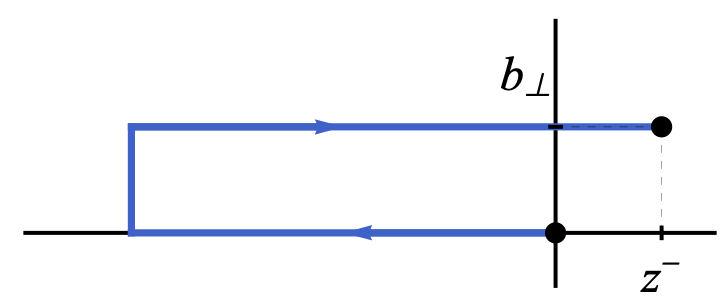}} 
    \caption{Wilson lines for (a) SIDIS process with the space-like correlation function and (b) Drell-Yan process with the timelike correlation.}
    \label{fig:SIDIS-DY}
\end{figure}

\section{Pion and kaon TMDs}
\label{SECII}
TMDs map the internal structure of a light-ray hadron, in terms of partons with fixed longitudinal momentum at a given separation in the transverse space. 
In this section, we will briefly review the general definition for the pion
unpolarized and polarized TMDs, and present arguments
for a soft approximation in the ILM at low resolution.

\subsection{General}
In the leading twist, the pion TMDs, or pion beam functions, are matrix elements of non-local quark bilinear operators,
with staple-shaped Wilson lines, 
\begin{widetext}
\bea
\label{tmd}
    q_\pi(x,k_\perp)&=&\int_{-\infty}^\infty\frac{dz^-}{2\pi}\int\frac{d^2b_\perp}{(2\pi)^2}e^{ix p^+z^--ik_\perp\cdot b_\perp}\langle \pi|\bar{\psi}(0) \gamma^+W^{(\pm)}[0,0_\perp; z^-,b_\perp]\psi(z^-,b_\perp)|\pi\rangle \nonumber\\
    \Delta q_\pi(x,k_\perp)&=&\int_{-\infty}^\infty\frac{dz^-}{2\pi}\int\frac{d^2b_\perp}{(2\pi)^2}e^{ix p^+z^--ik_\perp\cdot b_\perp}\langle \pi|\bar{\psi}(0) \gamma^+\gamma^5W^{(\pm)}[0,0_\perp; z^-,b_\perp]\psi(z^-,b_\perp)|\pi\rangle \nonumber\\
    \delta q_\pi(x,k_\perp)&=&\int_{-\infty}^\infty\frac{dz^-}{2\pi}\int\frac{d^2b_\perp}{(2\pi)^2}e^{ix p^+z^--ik_\perp\cdot b_\perp}\langle \pi|\bar{\psi}(0)i\sigma^{\perp+}\gamma^5 W^{(\pm)}[0,0_\perp; z^-,b_\perp]\psi(z^-,b_\perp)|\pi\rangle\nonumber\\
\eea
\end{widetext}
for the unpolarized, spin and transversity respectively. The transverse coordinate dependence can be obtained by using the Fourier transform
\bea
\tilde{q}_{\pi}(x,b_\perp)=\int\frac{d^2k_\perp}{(2\pi)^2}e^{ik_\perp\cdot b_\perp} q_{\pi}(x,k_\perp)
\eea
The pion is on-shell with energy $E_{\vec{p}}=\sqrt{m_\pi^2+\vec{p}^2}$ and $3$-momentum $\vec{p}$,
\begin{equation}
\begin{aligned}
    p^\mu=&p^+\bar n^\mu-\frac {m^2_\pi}{2p^+} n^\mu+p^\mu_\perp
\end{aligned}
\end{equation}
with hadron mass $m_\pi$. Here $n=(1,0,0_\perp)$ and $\bar n=(0,1,0_\perp)$ are the longitudinal and transverse  light cone vectors. Similar definitions are assumed for the kaons. 

\begin{figure*}
    \centering
    \includegraphics[width=1\linewidth]{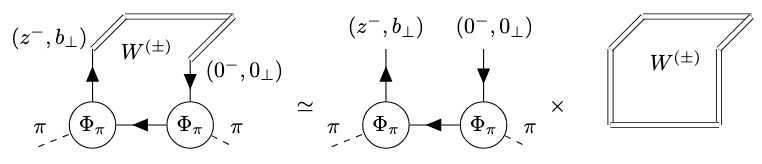}
    \caption{Soft separation for the quark pion TMD \eqref{tmd}, which is approximated into a constituent quark distribution without rapidity dependence and a rapidity dependent factor generated by the stapled Wilson line.}
    \label{fig:tmd_sub}
\end{figure*}

The Wilson line staples run space-like for SIDIS, and time-like for DY~\cite{Kumano:2020ijt}, as illustrated in Fig.~\ref{fig:SIDIS-DY}. 
The resummation of the collinear gluons from the final state in the SIDIS process, yields $W^{(+)}$. Similarly,  the resummation of the collinear gluons 
from the initial state in the Drell-Yan process, yields $W^{(-)}$~\cite{Angeles-Martinez:2015sea,GrossePerdekamp:2015xdx,Aidala:2012mv,Barone:2010zz,DAlesio:2007bjf}. More specifically, 
\begin{widetext}
\begin{equation}
\label{TMD_wilson}
    W^{(\pm)}[0,0_\perp; z^-,b_\perp] = W[0 ; \pm n\infty ] W[ \pm n\infty ; 
    \pm n\infty+b_\perp] W[ \pm n\infty+b_\perp ; nz^-+b_\perp].
\end{equation}
\end{widetext}
where the Wilson line from 
point $x$ to point $y$ reads 
\begin{equation}
    W[x;y]=\mathcal{P}\exp\left[ig\int_{x}^{y} dz_\mu A^\mu(z)\right]
\end{equation}
with a straight integration path. 
The stapled Wilson line in \eqref{TMD_wilson}, is composed of two Wilson lines along the light cone direction $n^\mu$, and one along the transverse direction. 


\subsection{$f,g,h$ TMDs}
In general, TMDs can be decomposed into several functions depending on the quark and target spin
orientations. The notational conventions are 
$f,g,h$ for unpolarized, longitudinally and transversely polarized TMDs, respectively. For the pion with spin-0,  there are only two TMDs
\bea
\label{pion_TMD}
    q_\pi(x,k_\perp)&=& f^{q/\pi}_1(x,k_\perp)\nonumber\\
    \delta q_\pi^\alpha(x,k_\perp)&=&
    \frac{\epsilon_\perp^{\alpha\beta} k^\beta_{\perp}}{m_\pi}h^{\perp,q/\pi}_{1}(x,k_\perp)
\eea
Here $\alpha, \beta$ are transverse and Cartesian indices, and the subscript 1 refers to the leading twist.


\subsection{Soft separation}
One important aspect of the pion TMDs in (\ref{tmd}) is the stapled Wilson contour. 
The standard factorization of the pion TMD at high resolution is briefly reviewed
in Appendix~\ref{App:tmd_fac}, involving soft subtraction to remove the rapidity (light-cone) singularity. In the ILM at low resolution, the gauge invariant
beam functions (\ref{tmd}) still suffer the rapidity divergence induced by the stapled Wilson line, 
where the cusp dependence from the instanton and anti-instanton
contributions are recently discussed in~\cite{Liu:2024sqj}. To capture the rapidity dependence, we will carry
its analysis  by assuming an 'approximate separation'  of the staple from the overlap of 
pion light cone wavefunction as illustrated in Fig.~\ref{fig:tmd_sub}, 
\begin{widetext}
\bea
\label{eq:pion_tmd_b}
    \tilde{f}^{q/\pi}_1(x,b_\perp,\mu\sim1/\rho,y_q-y_n)\approx \tilde{f}^{q/\pi}_{1}(x,b_\perp)e^{K^{(\rm inst)}_{\rm CS}(b_\perp/\rho)(y_q-y_n)}
\eea
\end{widetext}
This is justified for a large parton rapidity $y_q$ and only the cusp contribution is retained. 
The corresponding Collins-Soper (CS) kernel $K^{(\rm inst)}_{\rm CS}$ in the ILM, has been evaluated in~\cite{Liu:2024sqj}. This kernel introduces cusp dependence but is distinct from the perturbative gluon contribution to the CS kernel at high resolution~\cite{Ji:2004wu,Collins:2017oxh}. 
Here $\tilde{f}^{q/\pi}_{1}(x,b_\perp)$ on the right hand side is the low-resolution constituent quark distribution in pion calculated by the overlap of the pion light-front wave functions (LFWFs) in the ILM. The contribution from the stapled Wilson line is included in $e^{K^{(\rm inst)}_{\rm CS}(y_q-y_n)}$. The Wilson line rapidities and parton rapidities are defined as
\begin{align}
    &y_n=\frac12\ln\frac{n^+}{n^-} & &y_q=\frac12\ln\frac{k^+}{k^-}
\end{align}
where $k$ is the parton momentum.
The normalization of the unpolarized quark TMD is the standard PDF \cite{delRio:2024vvq,Gonzalez-Hernandez:2022ifv,Rogers:2020tfs}
\begin{equation}
    \int d^2k_\perp f^{q/\pi}_1(x,k_\perp)=f_1^{q/\pi}(x,\mu=1/\rho)
\end{equation}


\begin{figure*}
\centering
%
%
\subfloat[\label{fig:pion_contour}]{\includegraphics[width=0.325\linewidth]{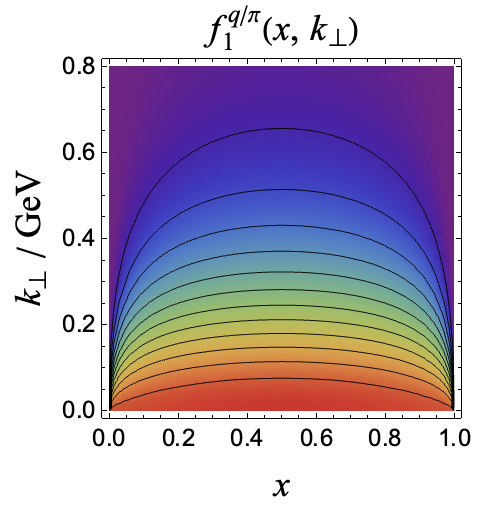}}
\hfill
\subfloat[\label{fig:kaon_contour}]{\includegraphics[width=0.325\linewidth]{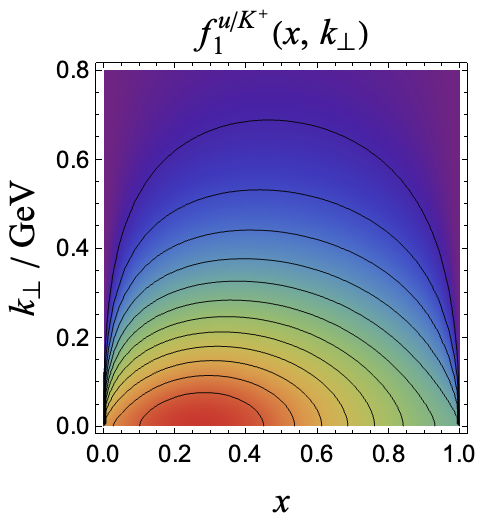}}
\hfill
\subfloat[\label{fig:pion_3d}]{\includegraphics[width=0.355\linewidth]{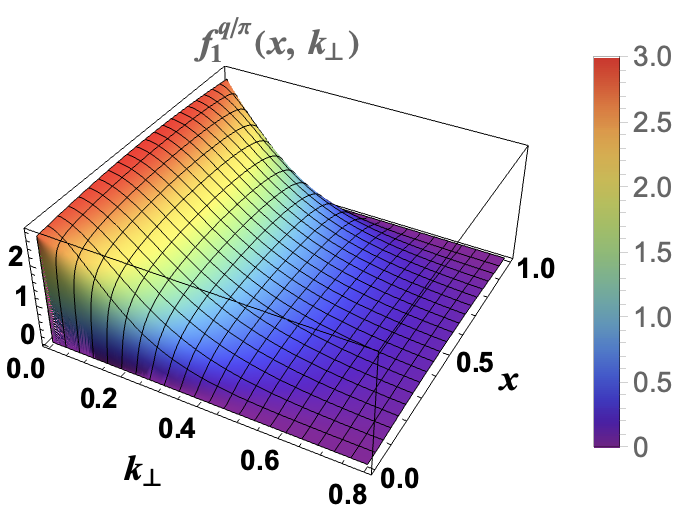}}
\hfill
\subfloat[\label{fig:kaon_3d}]{\includegraphics[width=0.355\linewidth]{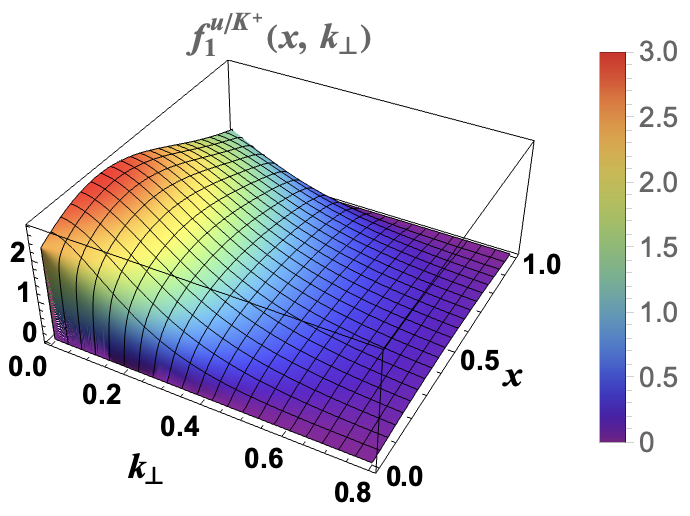}}
\hfill
\subfloat[\label{fig:pion_k}]{\includegraphics[width=0.38\linewidth]{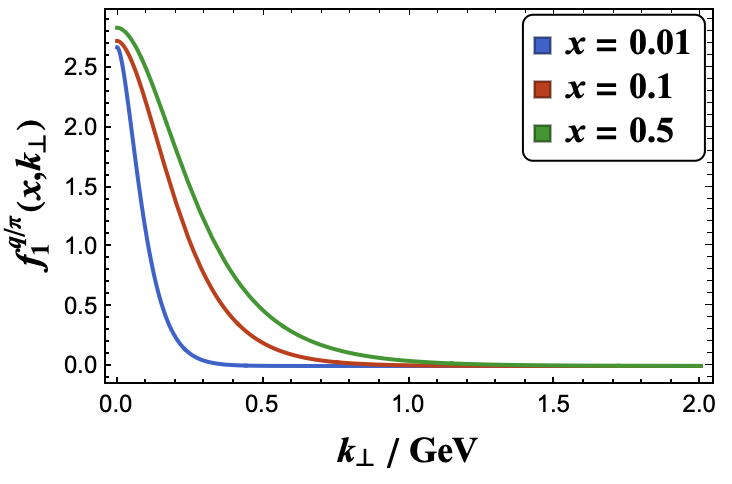}}
\hfill
\subfloat[\label{fig:pion_kk}]{\includegraphics[width=0.38\linewidth]{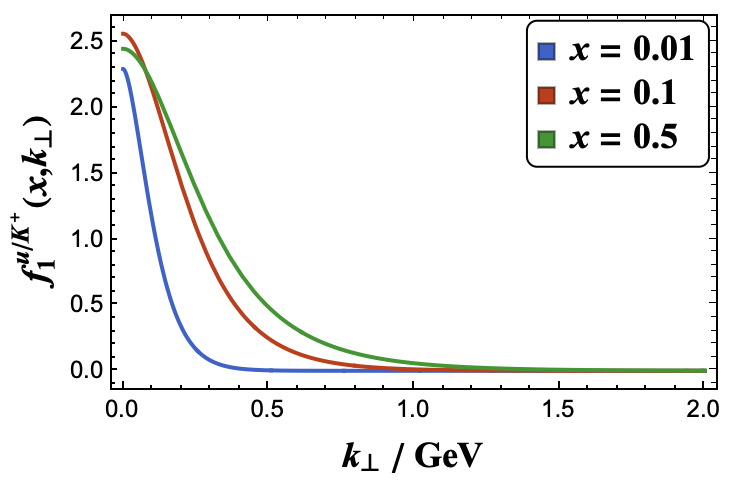}}
\hfill
\subfloat[\label{fig:pion_x}]{\includegraphics[width=0.38\linewidth]{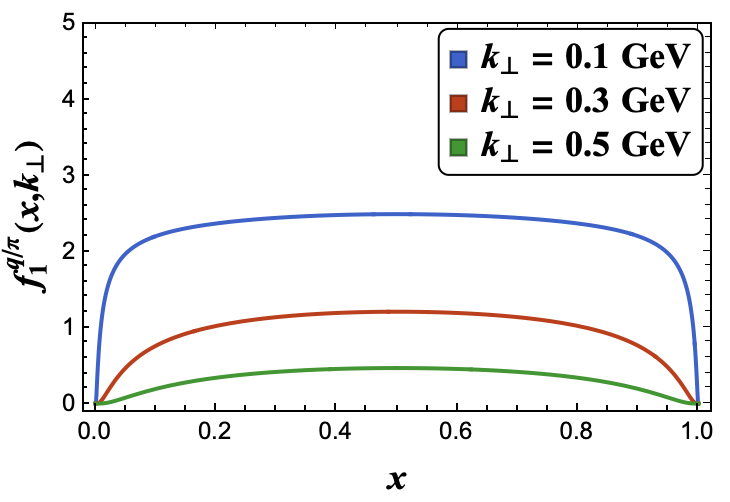}}
\hfill
\subfloat[\label{fig:pion_xk}]{\includegraphics[width=0.38\linewidth]{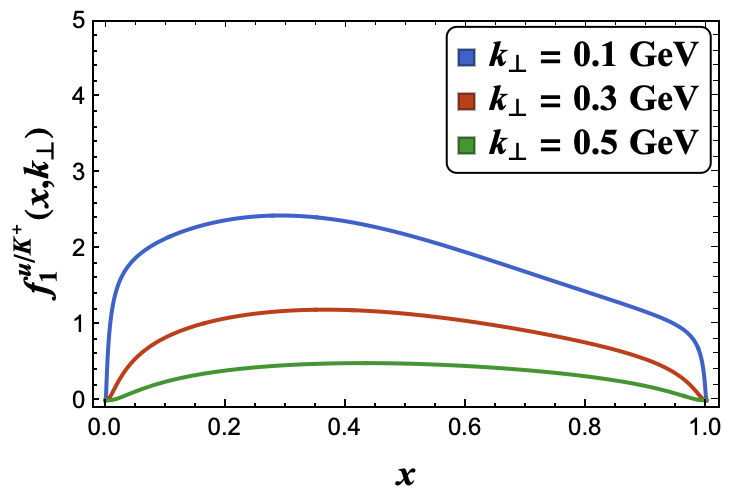}}
\caption{Constituent quark TMD distribution in pion (left) \eqref{eq:pion_tmd}  and kaon (right) 
\eqref{eq:kaon_tmd} at low resolution $\mu=1/\rho$
with $\rho=0.313$ fm: (a,b) are the density plots, (c,d) the 3D plots,
(e,f) the transverse momentum dependent plots for fixed $x$, and (g,h) the longitudinal momentum dependence for fixed $k_\perp$. The pion parameters are
$C_\pi=7.240$, $m_\pi=139.0$ MeV, $M=398.17$ MeV.
The kaon parameters are $C_K=6.60$, $m_K=458.0$ MeV, $M_u=394.4$ MeV, $M_s=556.5$ MeV.}
\label{fig:pion_TMD}
\end{figure*}

\subsection{Pion beam function}
Due to the soft separation at low resolution, the pion beam function \eqref{tmd} is divided into a constituent quark distribution and a Wilson staple factor carrying the cusp dependence. The constituent quark contributions in the pion, follow from the pion LFWFs in the lowest Fock space, 
\begin{widetext}
\bea
\label{FOCK1}
    q_\pi(x,k_\perp)&=&\frac{1}{(2\pi)^3}\sum_{s_1,s_2}\Phi^\dagger_\pi(x,k_\perp,s_1,s_2)\Phi_\pi(x,k_\perp,s_1,s_2)
\nonumber\\
    \Delta q_\pi(x,k_\perp)&=&\frac{1}{(2\pi)^3}\sum_{s_1,s_1',s_2}\Phi^\dagger_\pi(x,k_\perp,s_1,s_2)\sigma^3_{s_1s_1'}\Phi_\pi(x,k_\perp,s_1',s_2)
\nonumber\\
    \delta q_{\pi}(x,k_\perp)&=&\frac{1}{(2\pi)^3}\sum_{s_1,s_1',s_2}\Phi^\dagger_\pi(x,k_\perp,s_1,s_2)\sigma^{\perp}_{s_1s_1'}\Phi_\pi(x,k_\perp,s_1',s_2)
\eea
where $\sigma^{3,\perp}_{ss'}$ are the Pauli matrices and the pseudoscalar meson wave functions are defined as
\begin{equation}
\label{FOCK2}
    \Phi_{\pi}(x,k_\perp,s_1,s_2)
    =\frac{1}{\sqrt{N_c}}\left[\frac{C_{\pi}}{\sqrt{x\bar{x}}\left(m^2_\pi-\frac{k^2_\perp+M^2}{x\bar{x}}\right)}\mathcal{F}\left(\frac{k_\perp}{\lambda_\pi\sqrt{x\bar x}}\right)\right]\bar{u}_{s_1}(k_1)i\gamma^5\tau^a v_{s_2}(k_2)
\end{equation}
\end{widetext}
Here $u_{s_1}(k_1)$ refers to the light front quark spinor with spin $s_1$ and  internal momentum $k^+_1=xp^+$, while $k_{1\perp}=k_\perp$ and $v_{s_2}(k_2)$ refer to the light front anti-quark spinor with spin $s_2$ and  internal momentum $k^+_2=\bar{x}p^+$, $k_{2\perp}=-k_\perp$.

The instanton size induces the non-local form factor $\mathcal{F}(k)$ profiling the wave function with $\rho=0.313$ fm
\begin{equation}
\label{M_cut_off}
    \mathcal{F}(k)=\left[(zF'(z))^2\right]\bigg|_{z=\frac{k\rho}{2}}
\end{equation}
The normalization condition for the pion LFWF is
\begin{widetext}
\begin{equation}
    C_{\pi}=\left[\frac{1}{4\pi^2}\int_0^1dx\int_0^\infty dk^2_\perp\frac{k^2_\perp+M^2}{(x\bar{x}m_{\pi}^2-k_\perp^2-M^2)^2}\mathcal{F}^2\left(\frac{k_\perp}{\lambda_\pi\sqrt{x\bar x}}\right)\right]^{-1/2}
\end{equation}
\end{widetext}
with the parameter $\lambda_\pi$ fixed by the pion decay constant $f_\pi$, using the same LFWF contribution to the pion DA, see Appendix~\ref{App:LFWF}. More specifically, for the canonical ILM parameters, giving a constituent quark mass of $M=398.2$ MeV, a pion mass of
$m_\pi=139$ MeV,  and a pion decay constant of $f_\pi=94$ MeV, we have $\lambda_\pi=2.550$. More details regarding this construction can be found in~\cite{Liu:2023yuj,Liu:2023fpj}.

Inserting (\ref{FOCK2}) into (\ref{FOCK1}) and carrying the spin contraction, yield the constituent quark TMD distribution in a unpolarized pion 
\begin{equation}
\begin{aligned}
\label{eq:pion_tmd}
    &f^{q/\pi}_{1}(x,k_\perp)\\
    =&\frac{C^2_{\pi}}{(2\pi)^3}\frac{2(k_\perp^2+M^2)}{\left(x\bar{x}m^2_\pi-k^2_\perp-M^2\right)^2}\mathcal{F}^2\left(\frac{k_\perp}{\lambda_\pi\sqrt{x\bar x}}\right)
\end{aligned}
\end{equation}
and the null TMD pion Boer-Mulders function 
\begin{equation}
    h^{\perp,q/\pi}_{1}(x,k_\perp)=0
\end{equation}
as originally noted in~\cite{Lorce:2016ugb}. 
For simplicity, we will only focus on $f^{q/\pi}_1$ TMD PDF.
A non-zero  Boer-Mulders function, requires a cross contribution from the staple~\cite{Noguera:2015iia}.  T-odd TMDPDFs emerge from the gauge link structure, through a higher Fock contribution in the initial or final-state~\cite{JeffersonLabHallA:2011ayy,Kou:2023ady,Pasquini:2014ppa}. A similar approach using overlaps of LFWFs in the context of the spectator model to evaluate the Boer-Mulders function has been discussed in~\cite{Bacchetta:2008af}, and
applied to T-even gluon TMDs in~\cite{Bacchetta:2020vty} and T-odd gluon TMDs in~\cite{Bacchetta:2024fci}.

\subsection{Kaon beam function}
The constituent quark distribution in the kaon, follows from similar arguments. More specifically, the lowest Fock contribution to the kaon LFWF is
\begin{widetext}
\begin{equation}
    \Phi_{K}(x,k_\perp,s_1,s_2)
    =\frac{1}{\sqrt{N_c}}\left[\frac{C_{K}}{\sqrt{x\bar{x}}\left(m^2_K-\frac{k^2_\perp+\bar xM_u^2+ xM_s^2}{x\bar{x}}\right)}\mathcal{F}\left(\frac{k_\perp}{\lambda_K\sqrt{x\bar x}}\right)\right]\bar{u}_{s_1}(k_1)i\gamma^5\tau^a v_{s_2}(k_2)
\end{equation}
with the 1 assigned to the $u$-quark with constituent mass $M_u$, and  the 2 assigned to the $s$-quark with constituent mass $M_s$. The normalization condition fixes the normalization constant
\begin{equation}
    C_{K}=\left[\frac{1}{4\pi^2}\int_0^1dx\int_0^\infty dk^2_\perp\frac{k^2_\perp+\bar{x}^2M_u^2+x^2M_s^2+2x\bar xM_uM_s}{(x\bar{x}m_{K}^2-k_\perp^2-\bar xM_u^2- xM_s^2)^2}\mathcal{F}^2\left(\frac{k_\perp}{\lambda_K\sqrt{x\bar x}}\right)\right]^{-1/2}
\end{equation}
\end{widetext}
with $\lambda_K=3.069$ to reproduce  the kaon decay constant $f_K=112\,\rm MeV$. For our choice of the ILM parameters, the kaon mass is $m_K=458$ MeV, and the constituent quark masses are $M_u=394.4$ MeV and  $M_s=556.5$ MeV. A rerun of the preceding arguments for the kaon, yield the constituent quark TMD distribution in a unpolarized kaon. 
\begin{widetext}
\begin{equation}
\begin{aligned}
\label{eq:kaon_tmd}
    f^{u/K^+}_{1}(x,k_\perp)=\frac{C^2_{K}}{(2\pi)^3}\frac{2(k^2_\perp+\bar{x}^2M_u^2+x^2M_s^2+2x\bar xM_uM_s)}{(x\bar{x}m_{K}^2-k_\perp^2-\bar xM_u^2- xM_s^2)^2}\mathcal{F}^2\left(\frac{k_\perp}{\lambda_K\sqrt{x\bar x}}\right)
\end{aligned}
\end{equation}
\end{widetext}

The constituent quark TMD distributions in \eqref{eq:pion_tmd} and \eqref{eq:kaon_tmd}, defined through LFWFs with the stapled Wilson line softly separated in the instanton vacuum picture, are later associated with the soft subtracted quark TMD at low resolution in \eqref{pi_tmd}. 
In Fig.~\ref{fig:pion_TMD} we show constituent quark distribution evaluated by lowest Fock state LFWF overlap in pion (left) \eqref{eq:pion_tmd} and kaon (right) 
\eqref{eq:kaon_tmd} 
at low resolution $\mu=1/\rho$
with $\rho=0.313$ fm: Fig.~\ref{fig:pion_contour}, \ref{fig:kaon_contour} are the density plots, \ref{fig:pion_3d}, \ref{fig:kaon_3d} the 3D plots,
\ref{fig:pion_k}, \ref{fig:pion_kk} the transverse momentum dependent plots for fixed $x$, and \ref{fig:pion_x}, \ref{fig:pion_xk} the longitudinal momentum dependence for fixed $k_\perp$. The pion parameters are
$C_\pi=7.240$, $m_\pi=139.0$ MeV, $M=398.17$ MeV.
The kaon parameters are $C_K=6.60$, $m_K=458.0$ MeV, $M_u=394.4$ MeV, $M_s=556.5$ MeV.

\begin{figure*}
\centering
\subfloat[\label{fig:TMD_soft_a}]{\includegraphics[width=.45\linewidth]{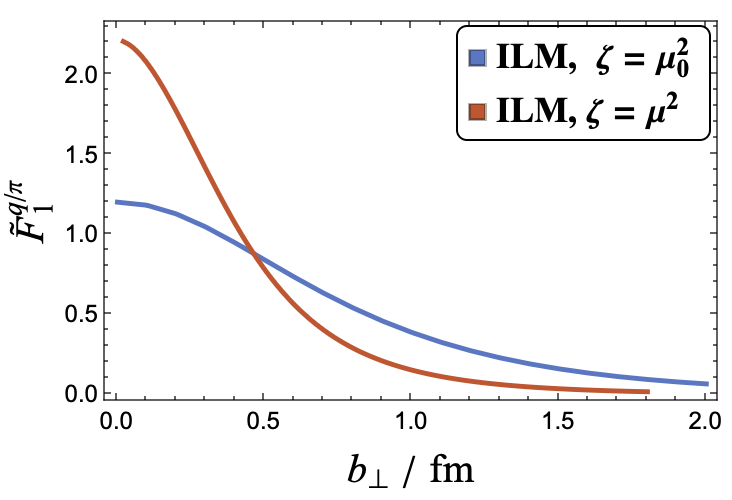}}
\hfill
\subfloat[\label{fig:TMD_soft_b}]{\includegraphics[width=.45\linewidth]{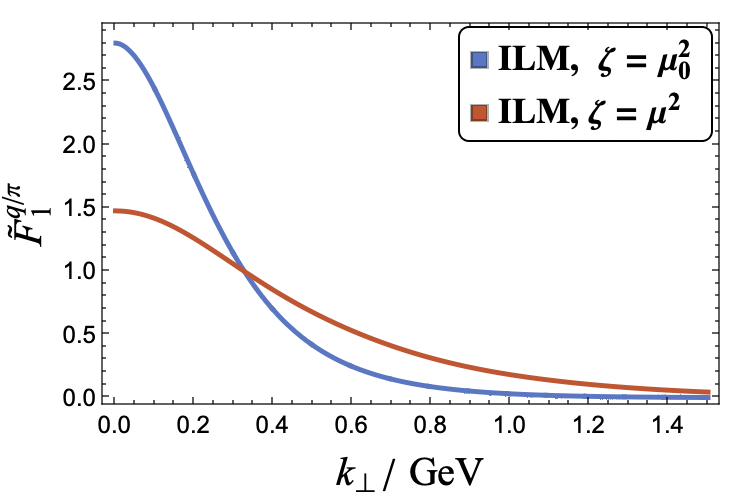}}
    \caption{The comparison of ILM estimated pion TMD distribution $\tilde{F}_1^{q/\pi}(x,b_\perp;\mu_0,\zeta)$ \eqref{pi_tmd} at scale $\mu_0\sim1/\rho$ (perturbative gluons excluded) between the one with full unsubtracted quark transverse momentum included (blue, $\zeta=\mu_0^2$) and soft transverse momentum subtracted (red, $\zeta=\mu^2=4$ GeV$^2$) at $x=0.3$ in $b_\perp$ space (a) and in $k_\perp$ space (b)}
    \label{fig:TMD_soft}
\end{figure*}

\section{TMD Evolution}
\label{SECIII}
The rapidity scale plays an important role in TMD evolution, as it distangles the soft exchanges between the collinear hadrons in phase space, in the context of TMD factorization. 
The rapidity divergences in \eqref{tmd} are controlled by Wilson line rapidity $y_n\rightarrow-\infty$.
With this in mind, the TMD beam functions are renormalized by subtracting the rapidity divergence with soft function $S$. The renormalized (soft-subtracted) TMD functions read~\cite{Ji:2004wu}.
\bea
&&\tilde{F}^{q/\pi}_1(x,b_\perp;\mu,\zeta)=\nonumber\\
&&\lim_{y_n\rightarrow-\infty}\frac{\tilde{f}^{q/\pi}_1(x,b_\perp,\mu,y_\pi-y_n)}{S(b_\perp,\mu,y-y_n)}
\eea
where $S$ is the soft factor given in \eqref{soft_factor}.
Here rapidity scale \cite{Collins:2011zzd}
$$\zeta=2(k^+)^2e^{-2y}
$$ 
represents the energy of a parton in the hadron, and  $y$ an arbitrary rapidity mark to 
account for the rapidity subtraction under evolution.

\begin{figure*}
    \centering
\subfloat[\label{tmd_b_a}]{\includegraphics[width=0.4\linewidth]{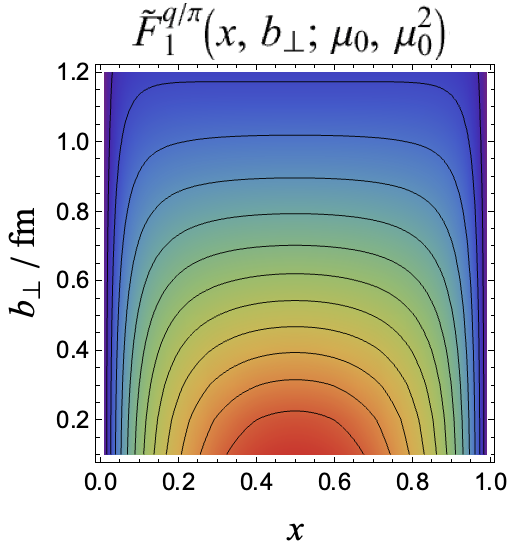}}
\hfill
\subfloat[\label{}]{\includegraphics[width=0.4\linewidth]{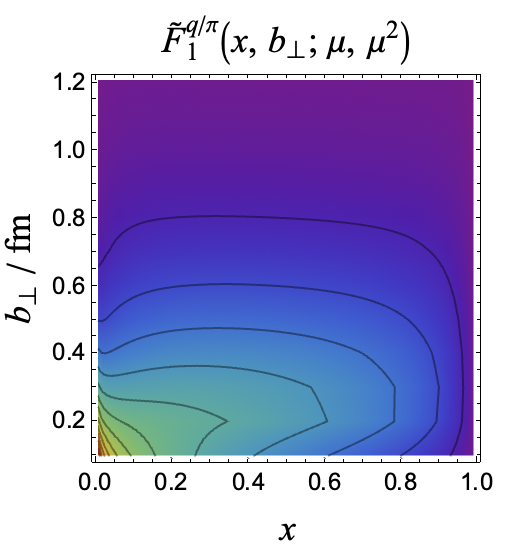}}
\hfill
\subfloat[\label{}]{\includegraphics[width=0.4\linewidth]{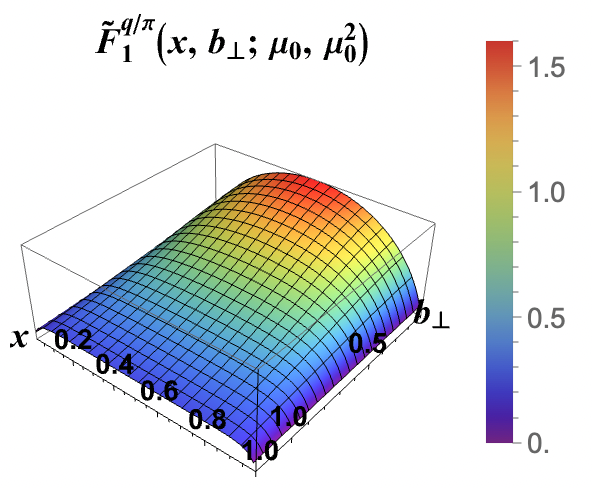}}
\hfill
\subfloat[\label{}]{\includegraphics[width=0.4\linewidth]{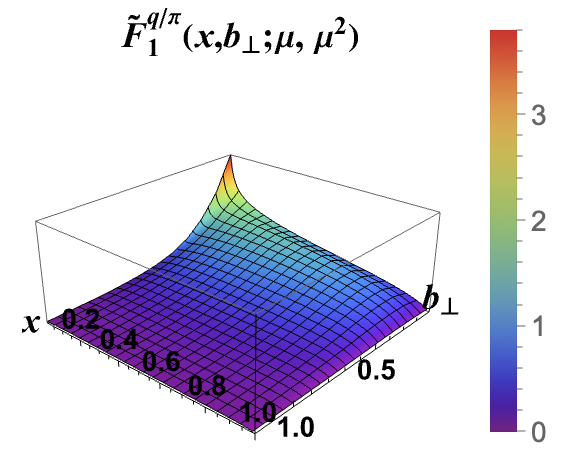}}
    \caption{Upper panel is the contour plot of the pion TMD parton distribution $\tilde{F}_1^{q/\pi}(x,b_\perp;\mu,\zeta)$: (a) soft subtracted quark TMD at intrinsic rapidity scale $\zeta=\mu_0^2$ and $\mu_0=1/\rho=0.63$ GeV 
    and (b) soft subtracted TMD evolved to $\mu=2$ GeV with chosen scheme $\zeta=\mu^2$. The lower panel is the 3D plot of the pion TMD parton distribution $\tilde{F}_1^{q/\pi}$: (c) quark TMD with the soft subtraction at intrinsic rapidity scale $\zeta=\mu_0^2$ and $\mu_0=1/\rho=0.63$ GeV
    and (d) subtracted TMD evolved to $\mu=2$ GeV. Here $b_\perp$ is in fm}
    \label{fig:tmd_b}
\end{figure*}

\begin{figure*}
    \centering
\subfloat[\label{}]{\includegraphics[width=0.4\linewidth]{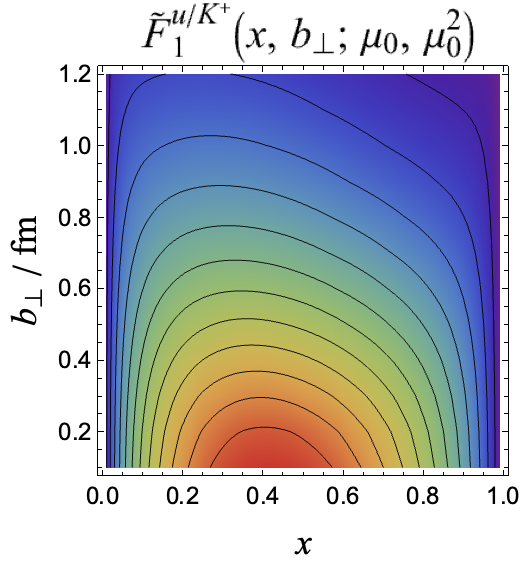}}
\hfill
\subfloat[\label{}]{\includegraphics[width=0.4\linewidth]{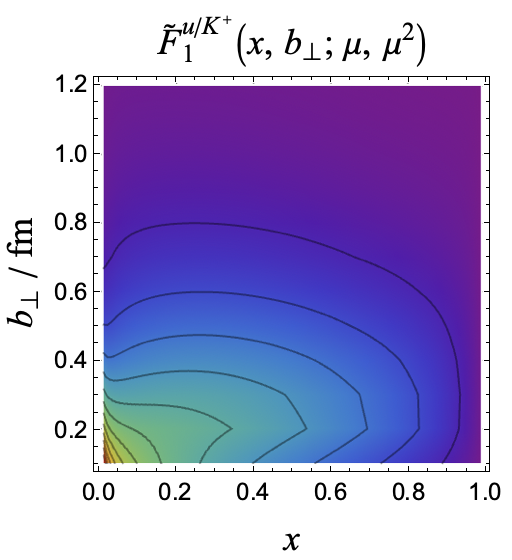}}
\hfill
\subfloat[\label{}]{\includegraphics[width=0.44\linewidth]{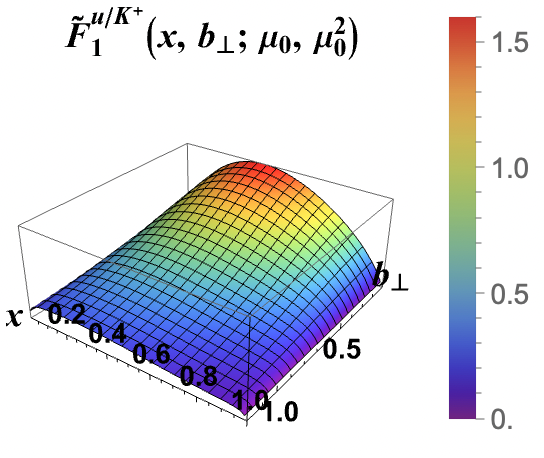}}
\hfill
\subfloat[\label{}]{\includegraphics[width=0.43\linewidth]{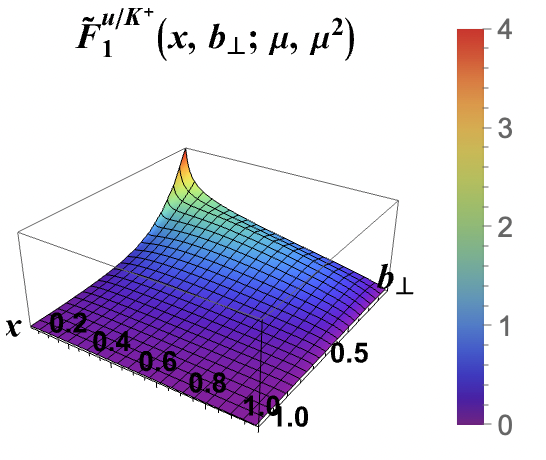}}
    \caption{Upper panel is the contour plot of the kaon TMD parton distribution $\tilde{F}_1^{u/K^+}$: (a) quark TMD with the soft subtraction at minimal rapidity scale $\zeta=\mu_0^2$ and $\mu_0=1/\rho=0.63$ GeV
    and (b) subtracted TMD  evolved to $\mu=2$ GeV. The lower panel is the 3D plot of the kaon TMD parton distribution $\tilde{F}_1^{u/K^+}$: (c) quark TMD with the soft subtraction at minimal rapidity scale $\zeta=\mu_0^2$ and $\mu_0=1/\rho=0.63$ GeV
    and (d) subtracted TMD evolved to $\mu=2$ GeV. Here $b_\perp$ is in fm}
    \label{fig:tmd_b_K}
\end{figure*}

The amount of subtraction is controlled by the  rapidity dependence $\zeta$. 
At low resolution $\mu_0$, the soft-subtracted TMD $\tilde{F}^{q/\pi}_1(x,b_\perp;\mu_0,\zeta_0)$ in the ILM, as given by \eqref{eq:pion_tmd_b}, is defined as
\begin{equation}
\begin{aligned}
\label{pi_tmd}
    \tilde{F}&_1^{q/\pi}(x,b_\perp;\mu_0,\zeta_0)=\\
    &\tilde{f}_{1}^{q/\pi}(x,b_\perp)e^{K^{(\rm inst)}_{\rm CS}(b_\perp/\rho)\ln\sqrt{\frac{\zeta_0}{1/\rho^2}}}
\end{aligned}
\end{equation}
where 
$\tilde{f}^{q/\pi}_{1}(x,b_\perp)$ is evaluated in \eqref{eq:pion_tmd} and \eqref{eq:kaon_tmd} for kaons.
%
%
The variation in the rapidity $y$ determines the $\zeta$ evolution of TMDs  \cite{Boer:2014tka,Idilbi:2004vb,Aybat:2011zv,Collins:2003fm}.
Generally, this  is implemented in the $b_\perp$ space Fourier-conjugate to $k_\perp$, allowing for a clearer separation of perturbative and non-perturbative effects~\cite{Collins:2011zzd}.
For any leading-twist TMD  with soft subtraction $\tilde{F}$, the CSS renormalization group equations reads \cite{Boer:2015ala,Collins:2014loa}
\begin{align}
\label{tmd_evol_rg}
    \frac{d}{d\ln\sqrt{\zeta}} \ln \tilde{F}(x, b_\perp; \mu,\zeta)=&K_{\rm CS}(b_\perp,\mu)\nonumber\\
    \frac{d}{d\ln\mu} \ln\tilde{F}(x, b_\perp; \mu,\zeta)=& \Gamma_F (\mu,\zeta)
\end{align}
where 
\begin{equation}
    \Gamma_F(\mu,\zeta)=\gamma_F (\alpha_s(\mu))\nonumber- \Gamma_{\mathrm{cusp}}(\alpha_s(\mu)) \ln\left(\frac{\zeta}{\mu^2}\right)
\end{equation}
The CS kernel $K_{\rm CS}$ drives
 the rapidity $\zeta$ scale evolution. The explicit forms of the anomalous dimensions $\gamma_F$,  depend on the quark operator insertion $\gamma^+,\gamma^+\gamma^5, i\sigma^{\alpha+}\gamma^5$. Their perturbative expansions are given in Appendix \ref{Appx:cusp}. The evolved  TMDs involve one intrinsic transverse scale $b_\perp$ and two evolution scales $\mu$, $\zeta$, both driven
 by perturbative and nonperturbative physics. 

A consistent matching between the TMD evolution in the non-perturbative part of the transverse space,  to the standard and perturbative collinear factorization part, is  notoriously subtle. For completeness, we note the recent development on the $k_\perp$ momentum space TMD approach, related to the non-perturbative TMD evolution in~\cite{Aslan:2024nqg,Gonzalez-Hernandez:2023iso,Gonzalez-Hernandez:2022ifv}. For a more thorough review, see \cite{Rogers:2024cci}. Alternatively, we use a simple  procedure in $b_\perp$ space, that consists in minimal  matching of the large and small $b_\perp$ regions.

 \begin{figure}
    \centering
    \includegraphics[width=0.8\linewidth]{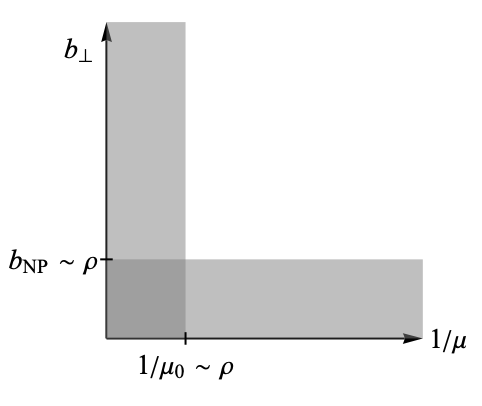}
    \caption{Perturbative (shaded) and non-perturbative (unshaded) regions
    for TMD evolution in transverse $b_\perp$ and resolution $\mu$.}
    \label{fig:scale}
\end{figure}
 
\begin{figure*}
    \centering
\subfloat[\label{fig:pion_pdf_1}]{\includegraphics[width=.5\linewidth]{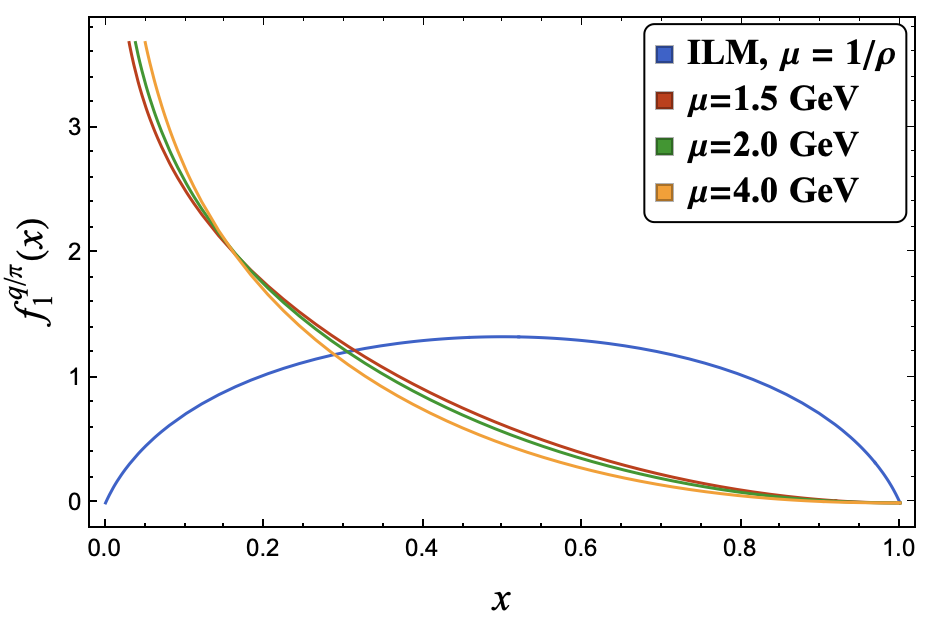}}
\hfill
\subfloat[\label{fig:pion_pdf_2}]{\includegraphics[width=.5\linewidth]{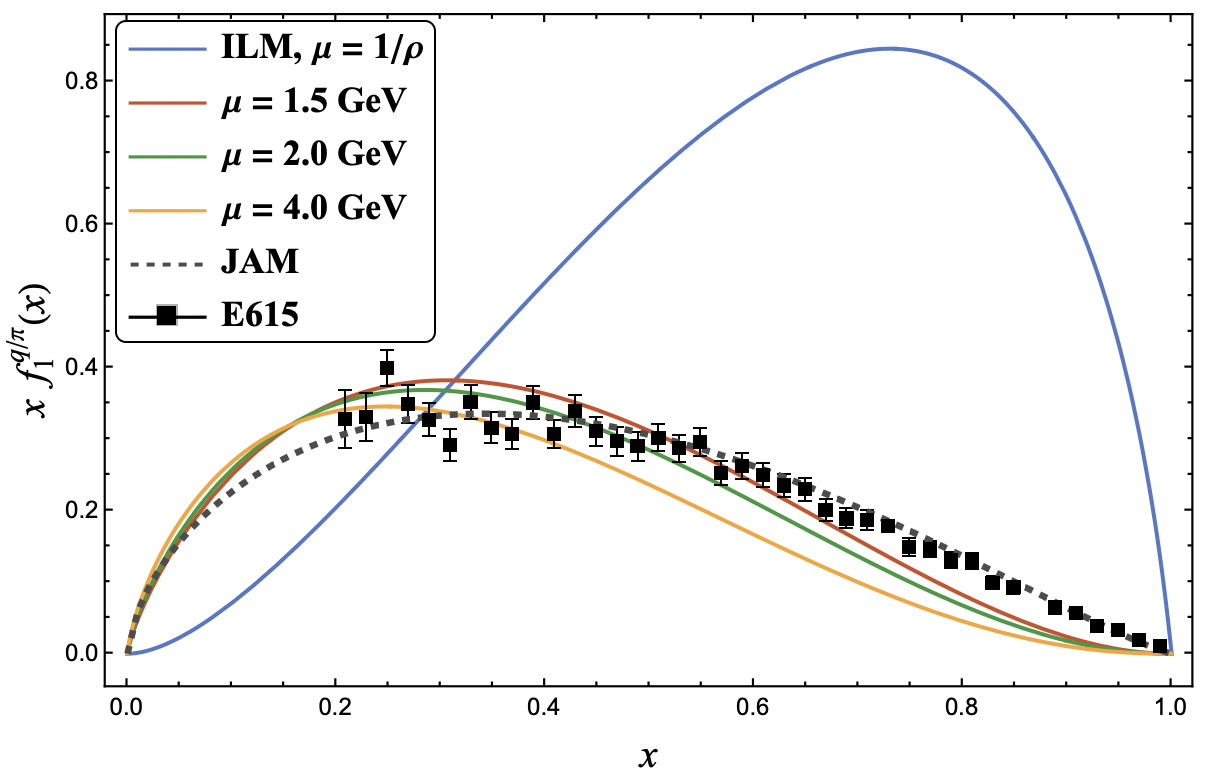}}
    \caption{ILM estimated parton density distribution (a) and momentum distribution (b) of pion evolved with Dokshitzer–Gribov–Lipatov–Altarelli–Parisi (DGLAP) equation. The results in the ILM are from~\cite{Liu:2023yuj,Liu:2023fpj}. The momentum distribution is compared to E615 experimental data \cite{Conway:1989fs} and JAM global QCD analysis \cite{Aicher:2010cb}.}
    \label{fig:pion_pdf}
\end{figure*}

 For simplicity, we can choose a simple renormalization scheme where the rapidity evolution follows $\zeta=\mu^2$. In Fig.~\ref{fig:scale} we illustrate the regime of scales pertinent to perturbative
 evolution in shade. 
In the perturbative regime, the perturbative gluons are dominant, while in the opposite regime the primordial gluon epoxy (hard glue) \cite{Shuryak:2021fsu} is dominant. More specifically, in the perturbative regime ($2e^{-\gamma_E}/\mu\lesssim b_\perp\ll b_{\rm NP}$), we have~\cite{Bacchetta:2017vzh,Bacchetta:2017gcc,Barry:2023qqh,Collins:2011zzd}
\begin{widetext}
\begin{equation}
\begin{aligned}
\label{convo}
    &\tilde{F}_1^{q/\pi}(x, b_\perp; \mu,\mu^2)\big|_{b_\perp\ll b_{\rm NP}}
    \rightarrow\sum_{i=q, \bar q, g}\int_x^1 \frac{dx'}{x'}C_{q/i}(x/x',b_\perp,\mu_b)f_1^{i/\pi} (x',\mu_b) e^{-S_{\rm sud}(\mu_b,\mu)}
\end{aligned}
\end{equation}
\end{widetext}
where the transverse scale $\mu_b$ is defined as 
\begin{equation}
\label{mu_b}
\mu_b = 2 e^{-\gamma_E}/b_\perp
\end{equation}
with $\gamma_E$ the Euler–Mascheroni constant. This formulation follows the CSS formalism in~\cite{Collins:2017oxh},
and is perturbatively meaningful only at small values of $b_\perp$, with a scale $\mu_b$ that is sufficiently larger than the Landau pole $\Lambda_{\rm QCD}$.
The convolution is the operator product expansion (OPE) which describes the small-$b_\perp$ behavior of the TMDs in terms of the collinear PDFs $f_1^{q/\pi}$ convoluted with the perturbative Wilson coefficients $C_{i/j}$ \cite{Scimemi:2017etj} defined in Appendix~\ref{App:evol}. The PDFs with the evolution are shown in Fig.~\ref{fig:pion_pdf}. The Sudakov factor contains the perturbative effects of soft gluon radiation, driving the evolution of the TMDs from $\mu_b$ to the UV renormalization scale $\mu$ perturbatively. The  Sudakov kernel $S_{\rm sud}$ is defined as
\begin{widetext}
\begin{equation}
\label{sudakov}
    S_{\rm sud}(\mu_b, \mu)\approx \int_{\mu_b}^{\mu} \frac{d\mu'}{\mu'} \left[ \Gamma_{\rm cusp}(\alpha_s(\mu'^2)) \ln \left( \frac{\mu^2}{\mu'^2} \right) - \gamma_F(\alpha_s(\mu'^2)) \right]-K^{(\rm pert)}_{\rm CS}(b_{\perp},\mu_b)\ln\left(\frac{\mu}{\mu_b}\right)
\end{equation}
\end{widetext}

We note that when $b_\perp\rightarrow 0$, the perturbative formulation \eqref{convo} is not valid, and should be matched to the fixed-order collinear calculation at $\mu$ following from the OPE,
\begin{equation}
\begin{aligned}
    &\tilde{F}_1^{q/\pi}(x, b_\perp; \mu,\mu^2)\big|_{b_\perp\rightarrow0}
    \rightarrow f_1^{q/\pi} (x,\mu)
\end{aligned}
\end{equation}
This can be improved by modifying the transverse scale \eqref{mu_b} in the Sudakov factor $S_{\rm sud}$ through a smooth substitution $\mu_b \rightarrow \mu$ as $b_\perp \rightarrow 0$,  so that the lower limit of the integral in \eqref{sudakov} never exceeds the upper limit. This is similar to the transverse scale introduced in \cite{Bacchetta:2017gcc,Bacchetta:2022awv,Rogers:2024cci,Boer:2014tka,Collins:2016hqq}. In a way, this modification corresponds to a re-summation of logarithms,  such that the Sudakov exponent $S_{\rm sud}$ interpolates to $0$ at $b_\perp = 0$, as it should~\cite{Parisi:1979se,Altarelli:1984pt}.

In the non-perturbative region ($b_\perp\gg b_{\rm NP}$) as illustrated by the unshaded region
in Fig.~\ref{fig:scale}, the glue contribution in the ILM starts to modify the perturbative prediction in \eqref{convo}. As the transverse distance grows, the soft (transverse) scale $\mu_b$ in the Sudakov factor saturates to $\mu_0$. A similar $b_*$ prescription with the transverse scale $\mu_b$ saturated at large $b_\perp$, has been discussed in~\cite{Bozzi:2010xn,Boer:2014tka}.  The perturbative gluon radiation at large transverse distance is suppressed, and the nonperturbative component of the TMD evolution compensates the gluon radiation. A phenomenological illustration of this mechanism is discussed in~\cite{Collins:2011zzd,Collins:2014jpa}, where the transverse scale is smoothly saturated at large distance $\mu_b(b_\perp\rightarrow\infty)\rightarrow\mu_0$, to ensure that the perturbative evolution remains within the perturbative regime. With this in mind, the soft-subtracted TMD in large distance can be defined as
\begin{widetext}
\begin{equation}
\begin{aligned}
\label{F_np}
    \tilde{F}_1^{q/\pi}(x,b_\perp;\mu,\mu^2)\big|_{b_\perp\gg b_{\rm NP}}&\simeq~\tilde{F}^{q/\pi}_1(x,b_\perp;\mu_0,\mu_0^2) e^{K^{(\rm inst)}_{\rm CS}(b_\perp/\rho)\ln\sqrt{\frac{\mu^2}{\mu^2_0}}}\\
    &\times \exp\left[\int_{\mu_0}^{\mu} \frac{d\mu'}{\mu'}\left(\gamma_F(\alpha_s(\mu'^2))-\Gamma_{\rm cusp}(\alpha_s(\mu'^2)) \ln \left( \frac{\mu_0^2}{\mu'^2} \right) \right)\right]
\end{aligned}
\end{equation}

\end{widetext}
The perturbative Sudakov form factor $S_{\rm sud}$ evolves from $\mu_0$ to $\mu$.
At the initial scale $\mu=\mu_0\sim1/\rho$, the large distance TMD can be fully estimated by the instanton vacuum. The amount of the soft subtraction is controlled by the rapidity dependence $\zeta$. 
The full TMD function at high resolution $\mu$ interpolates between the large $b_\perp$ and small $b_\perp$ by optimal matching with the smooth cut-off $b_{\rm NP}$ in linear combination.
\begin{widetext}
\bea
\label{INTER}
\tilde{F}_1^{q/h}(x, b_\perp; \mu,\mu^2)=c_{\mathrm{s}}(b_\perp,b_{\rm NP})\tilde{F}_1^{q/h}(x, b_\perp\ll b_{\rm NP}; \mu,\mu^2)
+c_{\mathrm{l}}(b_\perp,b_{\rm NP})\tilde{F}_1^{q/h}(x, b_\perp\gg b_{\rm NP}; \mu,\mu^2)
\nonumber\\
\eea 
\end{widetext}
Here  $b_{\rm NP}$ is an optimized cross-over with a value of about $\rho$, that interpolates 
the perturbative and non-perturbative contributions. The matching coefficient functions $(c_{\mathrm{s}}, c_{\mathrm{l}})$ can be parameterized by any smooth functions with the boundary condition where at short distance ($b_\perp\ll b_{\rm NP}$),  $(c_{\mathrm{s}}, c_{\mathrm{l}})\rightarrow (1, 0)$ and at long distance distance ($b_\perp\gg b_{\rm NP}$),  $(c_{\mathrm{s}}, c_{\mathrm{l}})\rightarrow (0, 1)$ and a cross-over occurs at $b_{\rm NP}$.



\begin{figure}
    \centering
    \includegraphics[width=1\linewidth]{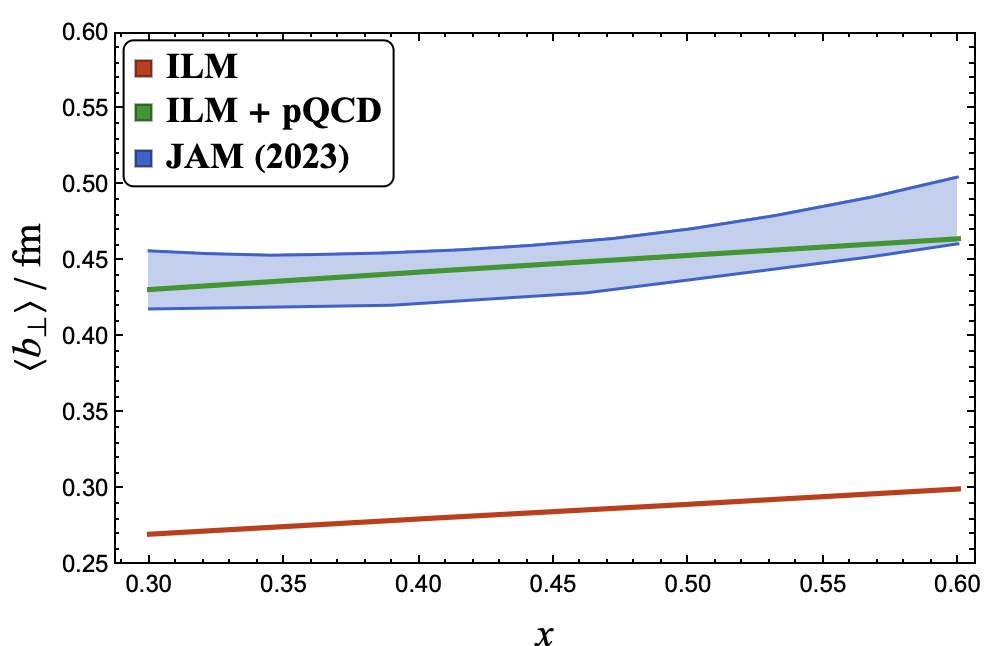}
    \caption{The instanton estimation on the conditional average of $b_\perp$ \eqref{eq:mean_b} is compared to the Drell-Yan experimental data extraction done by Jefferson Lab Angular Momentum (JAM) Collaboration~\cite{Barry:2023qqh} where the TMD evolution is taken from $\mu=1.27$ GeV to $\mu=4$ GeV. The red one is the estimation by ILM and the green curve is the RG improved result with N$^3$LO evolution kernel. The optimal matching cross-over $b_{\rm NP}$ in \eqref{INTER} is chosen to be $1.3$ GeV$^{-1}$}
    \label{fig:b_x}
\end{figure}

\begin{figure}
    \centering
\subfloat[\label{}]{\includegraphics[width=\linewidth]{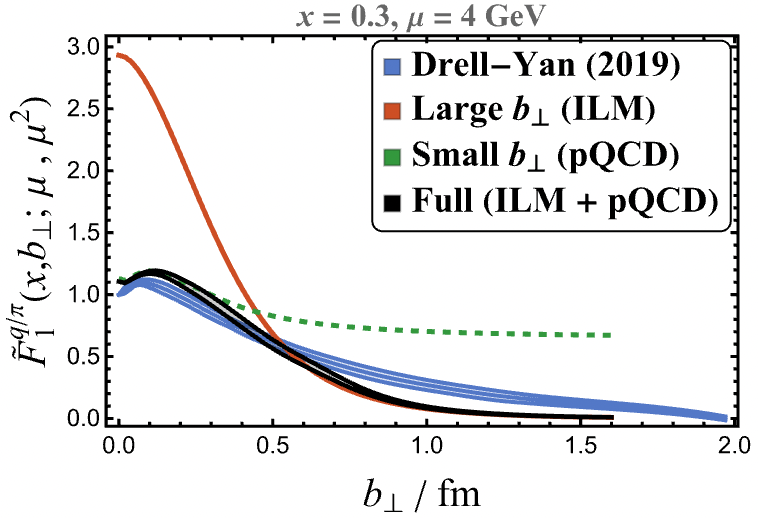}}
\hfill
\subfloat[\label{}]{\includegraphics[width=\linewidth]{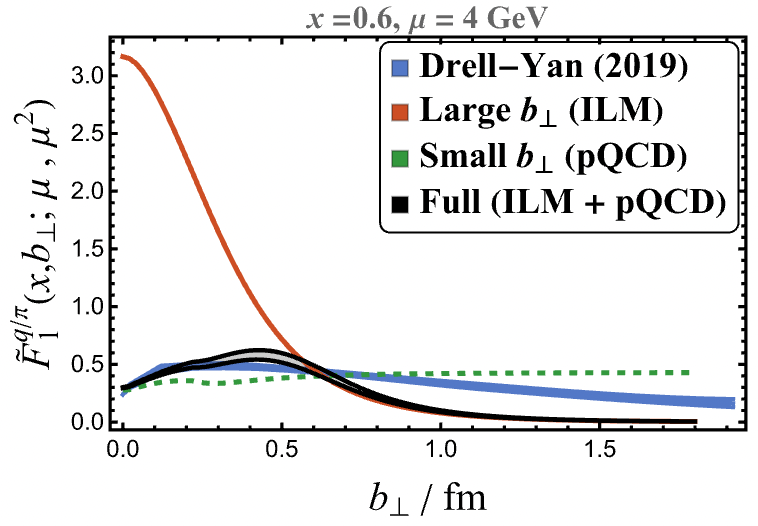}}
\caption{The evolved  pion TMD versus  $b_\perp$ with (a) $x=0.3$ and (b) $x=0.6$, compared to the Drell-Yan experimental data extraction \cite{Vladimirov:2019bfa} from three available measurements of the transverse momentum cross section for pion-induced Drell-Yan process performed by NA3 \cite{NA3:1982ntq}, E537 \cite{Anassontzis:1987hk} and E615 \cite{Conway:1989fs}.  The black band represents the optimized result with the propagated uncertainty from the original data.}
    \label{fig:f_b}
\end{figure}

\begin{figure*}
    \centering

\subfloat[\label{}]{\includegraphics[width=.48\linewidth]{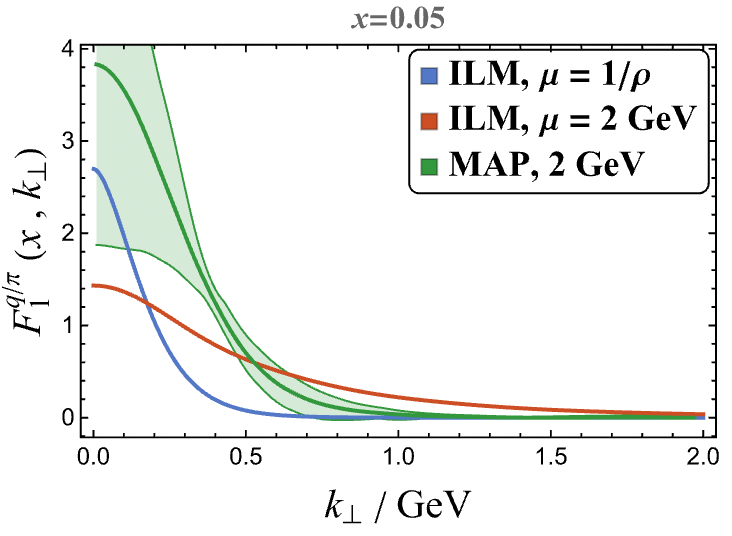}}
\hfill
\subfloat[\label{}]{\includegraphics[width=.5\linewidth]{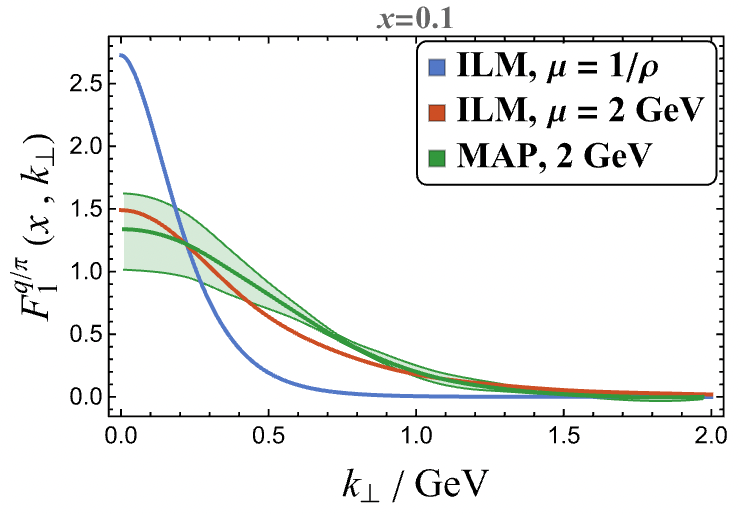}}
\hfill
\subfloat[\label{}]{\includegraphics[width=.5\linewidth]{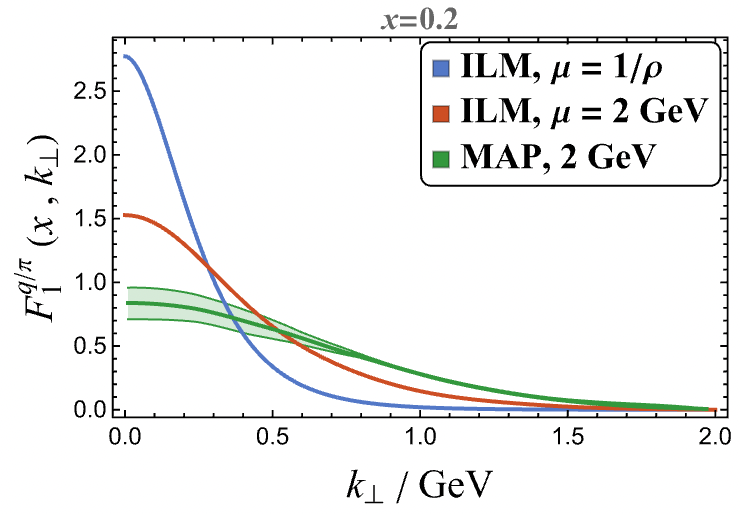}}
\hfill
\subfloat[\label{}]{\includegraphics[width=.5\linewidth]{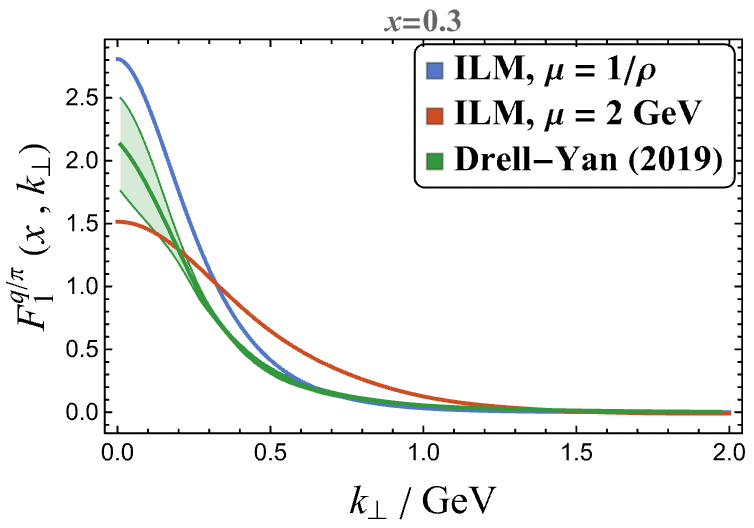}}
\caption{The evolved pion TMD versus $k_\perp$ (Fourier transform of \eqref{INTER}) with the optimized matching cross-over $b_{\rm NP}=0.26$ fm}, with (a) $x=0.05$, (b) $x=0.1$, (c) $x=0.2$, and (d) $x=0.3$. The results are compared to the extracted experimental data from the unpolarized pion-nucleus Drell–Yan process from \cite{Vladimirov:2019bfa} for $x=0.3$ and the MAP collaboration~\cite{Cerutti:2022lmb} for $x=0.05$-$0.2$.
    \label{fig:f_k}
\end{figure*}

\section{Results}
\label{SECIV}
In Fig.~\ref{fig:TMD_soft_a}, we show the pion TMD $\tilde{F}_1^{q/\pi}(x,b_\perp;\mu_0,\zeta)$ at the inverse instanton size resolution $\mu_0=1/\rho$ and
parton momentum fraction $x=0.3$, versus $b_\perp$  with minimal rapidity scale $\zeta=\mu_0^2$ (blue-solid curve) and higher rapidity $\zeta=\mu^2$ (red-solid curve). In Fig.~\ref{fig:TMD_soft_b}, we show the Fourier transform of the same results versus $k_\perp$.

In the upper panel of Fig.~\ref{fig:tmd_b} we show the contour plots for our final results for the pion TMDPDFs with $b_\perp$ measured in fm: (a)  shows the contour for the pion TMD   
$\tilde{F}_1^{q/\pi}$ at intrinsic parton rapidity $\zeta=\mu_0^2$ and $\mu_0=1/\rho=0.63$ GeV;  
(b) is the contour plot for the final subtracted pion TMD $\tilde{F}_1^{q/\pi}$ evolved to $\mu=\sqrt{\zeta}=2$ GeV with $b_{\rm NP}=1.3$ GeV$^{-1}$. The lower panel  of
Fig.~\ref{fig:tmd_b} is the 3D plots of 
the same results shown in the upper panel. It is evident that the extent of subtraction is more pronounced at small $x$, where the soft gluons are more populated.

In the upper panel of Fig.~\ref{fig:tmd_b_K} we show the contour plots for our final results for the kaon TMDPDFs, also  with $b_\perp$ measured in fm: (a)  shows the contour for the kaon TMD   
$\tilde{F}_1^{u/K^+}(x,b_\perp;\mu_0,\zeta)|_{\zeta=\mu^2_0}=\tilde{f}_1^{u/K^+}(x,b_\perp)$ at $\mu_0=1/\rho=0.63$ GeV;  
(b) is the contour plot for the subtracted kaon TMD $\tilde{F}_1^{u/K^+}$ evolved to $\mu=\sqrt{\zeta}=2$ GeV with $b_{\rm NP}=1.3$ GeV$^{-1}$. The lower panel  of
Fig.~\ref{fig:tmd_b_K} is the 3D plots of the same results shown in the upper panel. The huge change in the evolved TMD at high resolution, in comparison to those at low resolution, stems from the large modifications following from the DGLAP evolution of the PDFs shown in Fig.~\ref{fig:pion_pdf}.


The conditional distribution moment of $b_\perp$ is defined as
\begin{equation}
\label{eq:mean_b}
    \langle b^n_\perp\rangle(x)=\frac{\int d^2b_\perp b^n_\perp\tilde{F}_1^{q/\pi}(x, b_\perp; \mu,\mu^2)}{\int d^2b_\perp \tilde{F}_1^{q/\pi}(x, b_\perp; \mu,\mu^2)}
\end{equation}
In Fig.~\ref{fig:b_x} we show the ILM estimate for the  conditional average of $b_\perp$ following from 
\eqref{eq:mean_b} (red-solid curve),
and the RG improved result with N$^3$LO evolution kernel (green-solid curve)
with the choice $b_{\rm NP}=1.3\,\rm {GeV}^{-1}$ in \eqref{INTER}. The evolved result is
compared to the experimentally extracted
moment by the Jefferson Lab Angular Momentum Collaboration (JAM)~\cite{Barry:2023qqh} (blue-wide band). The empirical TMD uses the Drell-Yan data 
with evolution from $\mu=1.27$ GeV to $\mu=4$ GeV.

In Fig.~\ref{fig:f_b} we show the results for the pion TMD in $b_\perp$  in fm,  with (a) $x=0.3$ and (b) $x=0.6$ for the ILM (red-solid curve),
for pQCD (green-solid curve), the combined TMD with ILM plus pQCD (black-solid curve) with the cross-over between \eqref{convo} and \eqref{F_np} $b_{\rm NP}\approx0.55$ fm in \eqref{INTER}. The comparison is  to the Drell-Yan experimental data extraction~\cite{Vladimirov:2019bfa}
(blue-wide band) from three available measurements of the transverse momentum cross-section for pion-induced Drell-Yan processes performed by NA3 \cite{NA3:1982ntq}, E537 \cite{Anassontzis:1987hk} and E615 \cite{Conway:1989fs}.  Our results compare fairly to the empirical  Drell-Yan results, for the reported values of  $x$.


In Fig.~\ref{fig:f_k} we show the ILM  evolved pion TMD in momentum space $F_1^{q/\pi}(x,k_\perp)$ (Fourier transform of \eqref{INTER}) versus $k_\perp$, for  $x=0.05$ (a), $x=0.1$ (b), $x=0.2$ (c), and $x=0.3$ (d). The results are compared to the extracted experimental data from the unpolarized pion-nucleus Drell–Yan process from \cite{Vladimirov:2019bfa} for $x=0.3$ and from the MAP collaboration~\cite{Cerutti:2022lmb} for $x=0.05$-$0.2$. 
Our results compare fairly to the empirical MAP and Drell-Yan results at larger $x$. The deviation at smaller $x=0.05$ with the MAP results is expected, as the sea partons become more dominant. The extension of  the Fock space description at low resolution (say from 2 to 4), 
should enhance the sea quark contribution, and improve the comparison at very low $x$.

\section{Conclusions}
\label{SECV}
We have developed an analysis of the pion and kaon TMDs using
the ILM of the QCD vacuum. 
The leading twist TMDs
are pertinent expectation values of quark bilinears dressed by stapled Wilson lines,
evaluated in luminal pion and kaon states. They provide  a tomographic 
description of the the pion and kaon parton content,
in both parton fraction $x$ and transverse space $b_\perp$. 

Our analysis shows that the emerging tomographic pictures of both 
the pion and kaon TMDs $\tilde{F}^{q/h}_1$ at different resolutions, appear strikingly different.
In the ILM at low resolution $\mu_0\sim 1/\rho$ and large $b_\perp$, 
the TMDs carry partons mostly around $x\sim \frac 12$ within a range
$b_\perp\sim \rho$ of the order of the meson size. In contrast after
CSS evolution to higher resolution, the TMDs  are dominated by
partons at small $x\sim 0$ in a narrow region of $b_\perp\ll \rho$. 
In light of this, it may be important to revisit the reliability of the
CSS evolution starting from $\mu_0\sim 1/\rho$, by re-adapting it to the gradient
flow technique~\cite{L_scher_2013,Hasenfratz:2019hpg} which revealed the ILM 
in the deeply cooled regime~\cite{Leinweber:1999cw}.

Nevertheless, most effective models for TMDs of the pion and kaon do not take into account CSS 
evolution~\cite{dePaula:2023ver,Zhu:2023lst,Ahmady:2019yvo}, or at least do not address the source of the rapidity scale dependence~\cite{Kou:2023ady,Kaur:2019jfa}, so they are not readily amenable to comparison
the present results. Our comparison to the extracted TMDs from Drell-Yan
data are not very conclusive, as the latters appear to be dominated by 
perturbative contributions well into the non-perturbative regime. 
We would welcome more data analyses, and lattice simulations for
further comparison. The role of higher Fock components and
molecular configurations in the ILM, would be important to consider.

\vskip 1cm
{\bf Acknowledgments\,\,}
We thank Marco Radici and Ted Rogers for their comments on the first version of this manuscript. 
This work  is supported by the Office of Science, U.S. Department of Energy under Contract  No. DE-FG88ER40388. This research is also supported in part within the framework of the Quark-Gluon Tomography (QGT) Topical Collaboration, under contract no. DE-SC0023646.

\appendix

\section{LF conventions}

\label{Appx:LFspinor}
Throughout, we  use the Kogut-Soper conventions with the Weyl chiral basis for  the gamma matrices
\begin{eqnarray}
&\gamma^0=\begin{pmatrix}
0 & \mathds{1} \\
\mathds{1} & 0 \\
\end{pmatrix} ~\
&\gamma^{i}=\begin{pmatrix}
0 & \sigma^i \\
-\sigma^i  & 0 \\
\end{pmatrix}
\end{eqnarray}
The LF components are normalized to 
\begin{equation}
\gamma^\pm=\frac{\gamma^0\pm \gamma^3}{\sqrt{2}}
\end{equation}
with the LF projectors
\begin{equation}
\mathcal{P}_+=\frac{1}{2}\gamma^-\gamma^+=\begin{pmatrix}
1 & 0 & 0 & 0 \\
0 & 0 & 0 & 0 \\
0 & 0 & 0 & 0 \\
0 & 0 & 0 & 1 \\
\end{pmatrix}
\end{equation}
and $\mathcal{P}_-=1-\mathcal{P}_+$.

The LF fermionic field is composed of the plus (good) component  $\psi_+=\frac12\gamma^-\gamma^+\psi$
\begin{equation}
    \psi=\psi_++\frac{i}{\partial_-}\frac{\gamma^+}2\left(i\gamma_\perp\cdot \partial_\perp-M\right)\psi_+
\end{equation}
after eliminating  the minus (bad) component $\psi_-=\frac12\gamma^+\gamma^-\psi$

The spin-dependent wave functions denotes the spin states in the creation of a quark-anti-quark pair.  The spin wave functions for each channels can be computed using the LF spinors for the quarks and anti-quarks. The spinors are defined as 
\bea
&&u_s(k)=\frac{1}{\sqrt{\sqrt{2}k^+}}\left(\slashed{k}+M\right)\begin{pmatrix}\xi_s \\ \xi_s \end{pmatrix}\bigg|_{k^-=\frac{k^2_\perp+M^2}{2k^+}}
\nonumber\\
&&v_s(k)=\frac{1}{\sqrt{\sqrt{2}k^+}}\left(\slashed{k}+M\right)\begin{pmatrix}-\eta_s \\ \eta_s \end{pmatrix}\bigg|_{k^-=\frac{k^2_\perp+M^2}{2k^+}}\nonumber\\
\eea
respectively. Here 
 $\xi_s$ ($\eta_s$) refers to a quark 2-spinor (anti-quark 2-spinor) with a spin pointing in the $z$-direction, and $M$ the constituent quark mass. The anti-quark 2-spinor is related to the quark 2-spinor by charge conjugation $\eta_s=i\sigma_2\xi_s^*=-2s\xi_{-s}$

\section{Pion and kaon DAs}
\label{App:LFWF}
The pion and kaon distribution amplitudes (DAs) are defined as 
\begin{widetext}
\bea
\label{pion_DA}
\varphi_{\pi}(x)&=&
\frac  1{if_\pi}\int_{-\infty}^{+\infty} \frac{dz^-}{2\pi}e^{ixp^+ z^-}\langle0|\overline{d}(0)\frac1{\sqrt{2}}\gamma^+\gamma_5W[0;z^-]u(z^-)|\pi\rangle
\nonumber\\
\varphi_{K}(x)&=&
\frac  1{if_K}\int_{-\infty} ^{+\infty} \frac{dz^-}{2\pi}e^{ixp^+ z^-}\langle0|\overline{s}(0)\frac1{\sqrt{2}}\gamma^+\gamma_5W[0;z^-]u(z^-)|K\rangle
\eea
Their explicit forms  in the ILM read~\cite{Liu:2023fpj,Liu:2023yuj},
\bea
&&\varphi_{\pi}(x)=
\frac{2\sqrt{N_c} M}{f_\pi}\frac{C_\pi}{8\pi^2}\int_0^{\infty} dk^2_\perp  \frac{1}{k_\perp^2+M^2-x\bar{x}m^2_\pi} \mathcal{F}^2\bigg(\frac{k_\perp}{\lambda_\pi \sqrt{x\bar x}}\bigg)
\nonumber\\
&&\varphi_{K}(x)=
\frac{2\sqrt{N_c} (\bar{x}M_u+xM_s)}{f_K}\frac{C_K}{8\pi^2}\int_0^{\infty} dk^2_\perp  \frac{1}{k_\perp^2+\bar{x}M_u^2+xM_s^2-x\bar{x}m^2_K} \mathcal{F}^2\bigg(\frac{k_\perp}{\lambda_K \sqrt{x\bar x}}\bigg)
\eea
\end{widetext}
where $\bar x =1-x$. 
Both distribution amplitudes are properly normalized as
\begin{equation}
    \int_0^1 dx \varphi_{\pi,K}(x)=1
\end{equation}
The normalization condition is ensured by choosing the parameters $\lambda_{\pi,K}$ with given decay constant $f_{\pi}=94$ MeV and $f_K=122$ MeV at $\mu=1/\rho$.
\begin{align}
\label{lambda}
    \lambda_\pi=&2.550 & \lambda_K=&3.069
\end{align}

\section{TMD factorization}
\label{App:tmd_fac}
TMD factorization offers a framework to describe high-energy processes with momentum transverse much smaller than the characteristic hard scale of the interaction, in  Drell-Yan (DY), semi-inclusive deep inelastic scattering (SIDIS), and hadron production in lepton-pair annihilation. The scattering process is decomposed into a collinear factor or TMD distribution/fragmentation functions for the two hadrons, and soft factors capturing the long-range soft gluon exchange. A detailed review can be found in \cite{Scimemi:2019mlf}.



For DY processes, the naive TMD factorization for scattering involving hadrons $A$ onto $B$ factors into
\begin{equation}
\begin{aligned}
\label{tmd_fac}
    d\sigma\sim&\tilde{f}_A(x_A,b_\perp,y_A-y_n)\tilde{f}_B(x_B,b_\perp,y_{\bar n}-y_B)\\
    &\times S(b_\perp,y_n-y_{\bar n}) 
\end{aligned}
\end{equation}
with the impact Fourier transforms
\begin{equation}
\begin{aligned}
\label{tmd_unsub_1}
    &\tilde{f}_h(x_h,b_\perp,y_h-y_n)\\
    =&\int\frac{d^2k_\perp}{(2\pi)^2}e^{ik_\perp\cdot b_\perp} f_h(x,k_\perp,y_h-y_n)
\end{aligned}
\end{equation}
where $y_h=\frac12\ln(p^+/p^-)$ is the hadron rapidity and $y_n=\frac12\ln(n^+/n^-)$ is the rapidity for the near-light-cone direction $n$ in Wilson line. This direction usually refers to the moving direction of the other hadron.  
The soft function is defined as~\cite{Collins:2011zzd,Ji:2004wu,Liu:2024sqj}
\begin{equation}
\begin{aligned}
\label{soft_factor}
    &S(b_\perp,y_n-y_{\bar n})=\\
    &\frac1{N_c}\Tr\langle W_{\pm\bar{n}}(0)W^\dagger_{\pm n}(0)W_{\pm n}(b_\perp)W^\dagger_{\pm\bar{n}}(b_\perp)\rangle
\end{aligned}
\end{equation}
where the half-infinite Wilson lines, which represents the intermediate gluon resummation during the process, are specified by the direction $v$ and the initial point $x$, . 
\begin{equation}
\begin{aligned}
\label{wilson_line_1}
    W_v(x) =& W[v\infty + x; x ]\\
    =& \mathcal{P} \exp \left[ig \int_0^\infty dzv_\mu A^\mu(vz + x)\right]
\end{aligned}
\end{equation}

While naive factorization is 
straightforward, applying the definitions in \eqref{tmd_fac} often leads to ambiguous overlaps between collinear and soft kinematic regions, leading to unregulated light-cone divergences caused by $y_n\rightarrow-\infty$ and $y_{\bar n}\rightarrow\infty$. To disentangle their contributions, rapidity regulators are introduced, allowing the formula to be recast as \cite{Ji:2004wu}.
\begin{widetext}
\begin{equation}
\begin{aligned}
    &\lim_{\substack{y_n\rightarrow-\infty \\ y_{\bar n}\rightarrow\infty}}\tilde{f}_A(b_\perp,y_A-y_n)\tilde{f}_B(b_\perp,y_{\bar n}-y_B) S(b_\perp,y_n-y_{\bar n})\\
    &\rightarrow\frac{\tilde{f}_A(b_\perp,y_A-(-\infty))}{S(b_\perp,y_1-(-\infty))}\frac{\tilde{f}_B(x_B,b_\perp,\infty-y_B)}{S(b_\perp,\infty-y_2)}S(b_\perp,y_2-y_1)
\end{aligned}
\end{equation}
\end{widetext}
Here $y_1$ and $y_2$ are arbitrary rapidities introduced to subtract the overcounting soft region. For simplicity, we can assume $y_1=y_2$. The original definition of TMD distributions $\tilde{f}$ in \eqref{tmd_unsub_1} are called unsubtracted TMD distributions. After subtraction, the remaining soft factors can be absorbed into renormalized TMD distributions. Conventionally, the soft subtracted TMD distributions are defined as \cite{Aybat:2011zv,Collins:2003fm}
\begin{equation}
\begin{aligned}
&\tilde{F}_h(x,b_\perp;\mu,\zeta)=\tilde{f}_h(x,b_\perp,\mu,y-(-\infty))\\
&\times\sqrt{\frac{S(b_\perp,\mu,\infty-y)}{S(b_\perp,\mu,y-(-\infty))S(b_\perp,\mu,\infty-(-\infty))}}
\end{aligned}
\end{equation}
where $\zeta=2(xp^+)^2e^{-y}=m_h^2x^2e^{2(y_h-y)}$ represents the energy of the hadron with $y$ distinguishing the energy between two hadrons, $\zeta_A\zeta_B=(2x_Ax_Bp^+_Ap^-_B)^2=Q^4$.  

\section{TMD Evolution}
\label{App:evol}
The TMD renormalization group (RG) equations of leading-twist TMD functions in \eqref{tmd_evol_rg} are driven by the Collin-Soper kernel $K_{\rm CS}$ that governs the rapidity $\zeta$ scale evolution and the anomalous dimension $\gamma_F$ depending on the anomalous dimension of the TMD quark operators.

The naive solution for the RG equation of the TMD distributions reads \cite{Collins:2014loa}
\begin{widetext}
\begin{equation}
\begin{aligned}
    \tilde{F}_h(x,b_\perp;\mu,\zeta)=&\tilde{F}_h(x,b_\perp;\mu_0,\zeta_0)\\
    &\times\exp\int^{(\mu,\zeta)}_{(\mu_0,\zeta_0)}\left[\frac{d\mu}\mu\left(\gamma_F (\alpha_s(\mu))- \Gamma_{\mathrm{cusp}}(\alpha_s(\mu)) \ln\left(\frac{\zeta_0}{\mu^2}\right)\right)+\frac{d\zeta'}{\zeta'}\frac{K_{\rm CS}(b_\perp,\mu)}2\right]
\end{aligned}
\end{equation}

In the perturbative region where $b$ approaches zero, the TMD distribution reads \cite{Bacchetta:2017vzh,Bacchetta:2017gcc,Barry:2023qqh,Collins:2011zzd,Collins:2017oxh}

\begin{equation}
\begin{aligned}
\label{convo_1}
    &\tilde{F}_1^{q/h}(x, b_\perp; \mu,\mu^2)\big|_{b_\perp\rightarrow0}
    \rightarrow\sum_{i=q, \bar q, g}\int_x^1 \frac{dx'}{x'}C_{q/i}(x/x',b_\perp,\mu)f_1^{i/\pi} (x',\mu)
\end{aligned}
\end{equation}
\end{widetext}
Here we choose a simple scheme where the evolution follows $\zeta=\mu^2$. 
The convolution is the operator product expansion (OPE) which factorize the small-$b$ behavior of the TMDs into the collinear PDFs $f^{q/h}$ convoluted with perturbative Wilson coefficients $C_{i/j}$ at scale $\mu_b$. The Sudakov factor accounts for the remaining scale dependence, evolving the TMD distribution back to $\mu$ from $\mu_b$. The perturbative Wilson coefficient for quarks up to $\mathcal{O}(\alpha_s)$ is given by \cite{Scimemi:2017etj}
\begin{equation}
\begin{aligned}
&C_{q/q}\left(x, b_\perp, \mu_b \right)=\delta(1-x)\\
&+\frac{\alpha_s}{4\pi}C_F \left[ -\frac{\pi^2}{6} \delta(1 - x) + 2(1 - x) \right]
\end{aligned}
\end{equation}
where no flavor and quark-gluon mixing are considered. 

\section{CS kernel}
The evolution of CS kernel is driven by cusp anomalous dimension
\cite{Collins:2014jpa,Scimemi:2016ffw,DAlesio:2014mrz,Collins:1984kg,Rogers:2015sqa},  
\begin{equation}
\label{RG_K}
    \frac{dK_{\rm CS}(b_\perp,\mu) }{d\ln \mu^2}=-\Gamma_{\mathrm{cusp}}(\alpha_s(\mu))
\end{equation}
which is garenteed by integrability.
The implementation of TMD evolution equations, though based on pQCD, requires certain prescriptions to prevent extending the calculations to the non-perturbative region. In~\cite{Collins:2011zzd}, the Collins-Soper kernel is divided into a perturbative contribution plus  a non-perturbative left-over separated by  a smooth cut-off $b_*$, 
\begin{equation}
\begin{aligned}
\label{CS_kernel}
    &K_{\rm CS}(b_\perp,\mu)\\
    =&K^{(\rm pert)}_{\rm CS}(b_*, \mu_{b_\ast})-2\int_{\mu_{b_\ast}}^\mu \frac{d\mu'}{\mu'}\Gamma_{\rm cusp}+K^{(\rm np)}_{\rm CS}(b_\perp)
\end{aligned}
\end{equation}
Here $\mu_{b_\ast} = \frac{2 e^{-\gamma_E}}{b_\ast}$ and $b_\ast$ is the modified transverse distance suggested in~\cite{Collins:2011zzd},  such that $b_\ast(b_\perp\rightarrow\infty) \rightarrow b_{\rm NP}$ to ensure that the evolution remains within the perturbative regime.

The first and second part in \eqref{CS_kernel} are given by the perturbative renormalization group (RG) equation \eqref{RG_K},  with the evolution defined by the perturbative cusp anomalous dimension and its interaction constant with minimal logarithmic dependence. The coefficients of $K^{(\rm pert)}_{\rm CS}$ for each order of $\alpha_s$ can be found in Appendix~\ref{Appx:cusp}. The third part $K^{(\rm np)}_{\rm CS}(b_\perp)$ in \eqref{CS_kernel} is the non-perturbative correction to the Collins-Soper kernel, which is $\mu$-independent.

\begin{figure}
    \centering
    \includegraphics[width=1\linewidth]{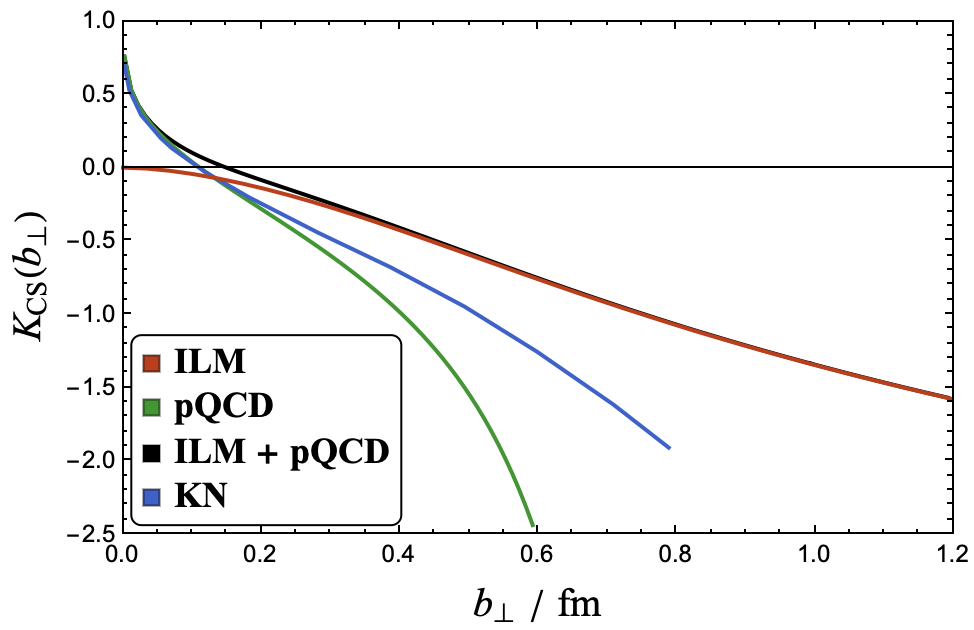}
    \caption{The instanton estimated Collin-Soper kernel (red curve) in \cite{Liu:2024sqj} and instanton estimation with RG improved by perturbative QCD evolution in \eqref{CS_kernel} (black curve) at small $b_\perp$ is compared to the pure perturbative RG (green curve) and data extraction (blue curve) from a global analysis of Drell-Yan lepton pair and $Z$ boson production \cite{Konychev:2005iy}}
    \label{fig:KCS}
\end{figure}

A simple model proposed in~\cite{Nadolsky:1999kb,Landry:2002ix,Konychev:2005iy} suggests
$K^{(\rm np)}_{\rm CS}$ to be a quadratic function, 
following the recent analysis in ~\cite{Bacchetta:2017vzh,Bacchetta:2017gcc}. 
In~\cite{Liu:2024sqj}, we have estimated the non-perturbative contribution in the ILM as
\begin{equation}
    K^{(\rm np)}_{\rm CS}(b_\perp)=K^{(\rm inst)}_{\rm CS}(b_\perp/\rho)-K^{(\rm inst)}_{\rm CS}(b_*/\rho)
\end{equation}
The CS kernel with the instanton contribution $K^{(\rm inst)}_{\rm CS}(b_\perp/\rho)$ is shown in Fig.~\ref{fig:KCS} (red curve) following from the analysis in~\cite{Liu:2024sqj}, suggesting a logarithmic dependence in large $b_\perp$ region. The non-perturbative parameter $b_{\rm NP}$ is chosen by matching the nonperturbative calculation (inst) with perturbative calculation (perp) at boundary $b_\perp=b_{\rm NP}$.
\begin{equation}
    K^{(\rm inst)}_{\rm CS}(b_{\rm NP}/\rho)=K^{(\rm perp)}_{\rm CS}(b_{\rm NP},\mu_{b_{\rm NP}})
\end{equation}

The renormalization group (RG) improved CS kernel with $b_{\rm NP}\sim0.13$ fm in \eqref{CS_kernel} (back-curve) is compared to the pure perturbative RG (green curve), and to the empirically extracted (blue curve) from a global analysis, from  Drell-Yan lepton pair and $Z$ boson production \cite{Konychev:2005iy}.


\section{Anomalous dimensions and Wilson coefficients for TMD evolution}
\label{Appx:cusp}
In this Appendix we summarise the anomalous dimensions and  Wilson coefficients for TMD evolution up to next-to-next-to-next-to-leading order (N$^3$LO)~\cite{Davies:1984hs,Collins:1984kg,Collins:2017oxh}

The  light-cone cusp anomalous dimension is given by
\begin{equation}
    \Gamma_{\mathrm{cusp}}(\alpha_s)=\sum_{n=1}^\infty 
    \left(\frac{\alpha_s}{4\pi}\right)^n\Gamma_n
\end{equation}
where $n$-th order coefficients at each order 
are defined as
\begin{equation}
    \Gamma_1=4C_F
\end{equation}

\begin{equation}
\begin{aligned}
    \Gamma_2 =&8 C_F \left[ C_A \left( \frac{67}{18} - \frac{\pi^2}{6} \right) - \frac{5}{9} N_f \right]\\
    =&\frac{1072}{9} - \frac{16 \pi^2}{3} - \frac{160}{27} N_f
\end{aligned}
\end{equation}

\begin{equation}
\begin{aligned}
    \Gamma_3 =& 352 \zeta(3) + \frac{176 \pi^4}{15} - \frac{2144 \pi^2}{9} + 1960 \\
    &+ N_f \left( -\frac{832 \zeta(3)}{9} + \frac{320 \pi^2}{27} - \frac{5104}{27} \right) \\
&- \frac{64}{81} N_f^2
\end{aligned}
\end{equation}

\begin{widetext}

\begin{equation}
\begin{aligned}
    \Gamma_4 =&\, 
    28032 \zeta(3)-1536 \zeta(3)^2 - 704 \pi^2 \zeta(3) - \frac{34496 \zeta(5)}{3} 
    + \frac{337112}{9} - \frac{178240 \pi^2}{27} + \frac{3608 \pi^4}{5} - \frac{32528 \pi^6}{945}\\
&+ N_f \Bigg( 
    \frac{1664 \pi^2 \zeta(3)}{9} - \frac{616640 \zeta(3)}{81} + \frac{25472 \zeta(5)}{9} 
 - \frac{1377380}{243} + \frac{51680 \pi^2}{81} - \frac{2464 \pi^4}{135} 
\Bigg)\\
&+ N_f^2 \Bigg( 
    \frac{16640 \zeta(3)}{81} + \frac{71500}{729} - \frac{1216 \pi^2}{243} - \frac{416 \pi^4}{405} 
\Bigg)\\
&+ N_f^3 \Bigg( 
    \frac{256 \zeta(3)}{81} - \frac{128}{243} 
\Bigg)\\
\end{aligned}
\end{equation}
\end{widetext}
The collinear anomalous dimension for the TMD operator is given by
\begin{equation}
    \gamma_{F}(\alpha_s)=\sum_{n=1}^\infty 
    \left(\frac{\alpha_s}{4\pi}\right)^n\gamma_n
\end{equation}
where the $n$-th order coefficients at each order are defined as
\begin{equation}
    \gamma_1 = 6 C_F
\end{equation}

\begin{equation}
\begin{aligned}
    \gamma_2 =&C_F^2 \left( 3 - 4 \pi^2 + 48 \zeta(3) \right)\\
    &+ C_F C_A \left( \frac{961}{27} + \frac{11 \pi^2}{3} - 52 \zeta(3) \right)\\
    &+ C_F N_f \left( - \frac{130}{27} - \frac{2 \pi^2}{3} \right)
\end{aligned}
\end{equation}

\begin{widetext}

\begin{equation}
\begin{aligned}
\gamma_3 =& C_F^2 N_f \left( - \frac{2953}{27} + \frac{26 \pi^2}{9} + \frac{28 \pi^4}{27} - 512 \zeta(3) \right) + C_F N_f^2 \left( - \frac{4834}{729} + \frac{20 \pi^2}{27} + \frac{16 \zeta(3)}{27} \right) \\
&+ C_F^3 \left( 29 + 6 \pi^2 + \frac{16 \pi^4}{5} + 136 \zeta(3) - \frac{32 \pi^2 \zeta(3)}{3} - 480 \zeta(5) \right) \\
&+ C_F^2 C_A \left( \frac{139345}{1458} + \frac{7163 \pi^2}{243} + \frac{83 \pi^4}{27} - \frac{7052 \zeta(3)}{9} + \frac{88 \pi^2 \zeta(3)}{9} + 272 \zeta(5) \right) \\
&+ C_A C_F N_f \left( \frac{17318}{729} - \frac{2594 \pi^2}{243} - \frac{22 \pi^4}{27} + \frac{1928 \zeta(3)}{27} \right) \\
&+ C_A C_F^2 \left( \frac{151}{2} - \frac{410 \pi^2}{9} - \frac{494 \pi^4}{135} + \frac{1688 \zeta(3)}{3} + \frac{16 \pi^2 \zeta(3)}{3} + 240 \zeta(5) \right)
\end{aligned}
\end{equation}
\end{widetext}

The 2D evolution of the transverse momentum dependence are driven by the anomalous dimension $\gamma_F$ and $\Gamma_{\rm cusp}$. They are identical between TMD PDFs and TMD fragmentation functions to all orders \cite{Collins:2017oxh}. The anomalous dimensions $\gamma_F$ only vary by the quark operator defining the TMD functions.

The perturbative part of the Collins-Soper kernel at minimal logarithmic scale $K^{(\rm pert)}_{\rm CS}(b_\perp, \mu=\mu_b)$ is defined as
\begin{equation}
    K^{(\rm pert)}_{\rm CS}(b_\perp, \mu_b)=\sum_{n=1}^\infty\left(\frac{\alpha_s(\mu_b)}{4\pi}\right)^nk_n
\end{equation}

\begin{align}
k_1 =& 0 \\
k_2 =&  8 C_F \left[\left( \frac{7}{2} \zeta(3) - \frac{101}{27} \right) C_A + \frac{14}{27} N_f \right]\nonumber\\
=&-2(-56 \zeta(3) + \frac{1616}{27} - \frac{224}{81} N_f)
\end{align}

\begin{widetext}
\begin{equation}
\begin{aligned}
k_3 =& -2\Bigg( \frac{176 \pi^2 \zeta(3)}{3} - \frac{24656 \zeta(3)}{9} + 1152 \zeta(5) + \frac{594058}{243} - \frac{6392 \pi^2}{81} - \frac{154 \pi^4}{45} \Bigg)  \\
&-2 N_f \left( \frac{7856 \zeta(3)}{81} - \frac{166316}{729} + \frac{824 \pi^2}{243} + \frac{4 \pi^4}{405} \right) -2 N_f^2 \left( \frac{64 \zeta(3)}{27} + \frac{3712}{2187} \right)
\end{aligned}
\end{equation}
\end{widetext}

\bibliography{ref}

\end{document}